\documentclass[aps,nofootinbib,twocolumn]{revtex4}
\usepackage{amsmath}
\usepackage{graphics}
\usepackage{amssymb}
\usepackage{multirow}
\usepackage{color}
\usepackage{xcolor}
\usepackage{graphicx}

\newtheorem{theorem}{Theorem}
\newtheorem{lemma}{Lemma}
\newtheorem{observation}{{\bf Observation}}

\newtheorem{definition}{Definition}
\newtheorem{corollary}{Corollary}

\def\be{\begin{equation}}
\def\ee{\end{equation}}
\def\ben{\begin{eqnarray}}
\def\een{\end{eqnarray}}

\begin{document}

\title{On the non-locality of tripartite non-signaling boxes emerging from wiring}

\begin{abstract}

It has been recently shown, that some of the tripartite boxes admitting bilocal decomposition,
lead to non-locality under wiring operation applied to two of the subsystems [R. Gallego {\it et al.} Physical Review Letters {\bf 109}, 070401 (2012)]. In the following, we study this phenomenon quantitatively. Basing on the known classes of boxes closed under wirings, we introduce multipartite monotones which are counterparts of bipartite ones - the non-locality cost and robustness of non-locality. We then provide analytical lower bounds on both the monotones in terms of the Maximal Non-locality which can be obtained by Wirings (MWN). We prove also upper bounds for the MWN of a given box, based on the weight of boxes signaling in a particular direction, that appear in its fully bilocal decomposition. We study different classes of partially local boxes (i.e. having local variable model with respect to some grouping of the parties). For each class the MWN is found, using the Linear Programming. The wirings which lead to the MWN and exhibit that some of them can serve as a witness of the certain classes are also identified. We conclude with example of partially local boxes being analogue of quantum states that allow to distribute entanglement in separable manner.


\end{abstract}

\author{Jan Tuziemski$^{1,3}$ and Karol Horodecki$^{2,3}$}
\affiliation{$^1$Faculty of Applied Physics and Mathematics,
Gda\'nsk University of Technology, 80--233 Gda\'nsk, Poland}
\affiliation{$^2$Institute of Informatics, University of Gda\'nsk, 80--952 Gda\'nsk,Poland}
\affiliation{$^3$National Quantum Information Centre of Gda\'nsk, 81--824 Sopot, Poland}

\date{\today}

\maketitle

\pagenumbering{arabic}

\section{Introduction}
The non-locality is one of the most intriguing characteristics of the quantum theory. Since seminal papers by Bell \cite{Bell}, where non-locality was referred as the non-local causality, as well as by Popescu and Rohrlich \cite{PR}, it has been treated as a resource \cite{Barret-Roberts} for the tasks such as communication complexity \cite{BruknerZPZ2002-Bell-complexity}, device independent cryptography \cite{BHK_Bell_key,MasanesPironioAcin,Bell-security,hanggi-2009,Hanggi-phd} or estimation of some properties of the system, like dimension \cite{BPAGMS-dim-witness} (see \cite{Bell-review} for a recent review). 

The central notion considered in the context of the non-locality is a conditional probability distribution, called a box. In the bipartite setting, the box determines the probability of obtaining results a and b, provided that measurement settings x and y were chosen. We are interested in the non-signaling boxes for which the change of measurement of one part does not change statistics of the other part. The box is (casually) local if it can be written as:
\ben
P(ab|xy) = \int_{\Lambda}  d \lambda q(\lambda) P(a|x,\lambda) P(b|y,\lambda),
\een
where $q(\lambda)$ is a distribution of a hidden variable $\lambda$ \cite{Bell-review}.
While the bipartite boxes have been studied deeply in recent years, the multipartite ones still deserve much attention. Due to the complicated and rich structure of the multipartite correlations some interesting results concerning the multipartite non-locality have been recently presented, opening an area for a further investigation. The conventional definition of the multipartite non-locality, due to Svetlichny \cite{Svetlichny}, states that if $P(a_1,a_2,a_3|x_1,x_2,x_3)$  can be written in the following way:
\ben
\label{eq-svd}
&&P(a_1,a_2,a_3|x_1,x_2,x_3) = \\ \nonumber &&\sum_{\lambda} p_\lambda  P_{\lambda}(a_1|x_1)P_{\lambda}(a_2,a_3|x_2,x_3) + \\ \nonumber &&\ \sum_{\mu} p_\mu  P_{\mu}(a_2|x_2)P_{\mu}(a_1,a_3|x_1,x_3) + \\ \nonumber &&\sum_{\nu} p_\nu  P_{\nu}(a_3|x_3)P_{\nu}(a_1,a_2|x_1,x_2),
\een
where $\sum_{\lambda} p_\lambda + \sum_{\mu} p_\mu + \sum_{\nu} p_\nu = 1$ and $\forall_\lambda p_\lambda \geq 0$, $\forall_\mu p_\mu \geq 0$, $\forall_\nu p_\nu \geq 0$, then it does not contain any tripartite non-locality, namely, it is local. In this paper the boxes admitting decomposition (\ref{eq-svd}) would be called the boxes with bilocal decomposition. It has been found that this definition has serious drawbacks \cite{BBGP2011-definition,GWAN2012-framework}. Namely, some of the boxes that are {\it local} according to this definition can entile signaling bipartite boxes in the decomposition (\ref{eq-svd}) which may lead to the so called grandfather type paradoxes \cite{BBGP2011-definition}. In turn, a new definition of multipartite (non-)locality has been proposed which eliminates the paradox. In parallel \cite{GWAN2012-framework}, another problem with the original definition has been found. Namely, when some of the parties that have an access to a multipartite box form a group, then they can create non-locality between the group as well as the rest of the parties by applying some processing of inputs and outputs called a {\it wiring} \cite{ABLPSV2009-closed-sets}. To avoid this phenomenon, which should not occur in case of {\it local} boxes, regardless what is their definition, an {\it operational framework} has been developed, as well as a new definition of multipartite non-locality has been proposed \cite{GWAN2012-framework}.

The both concepts of wiring and classes of non-local correlations have confirmed independently to be important in the context of non-locality. The wiring applied to many copies of a bipartite box, allow for an amplification of the weak correlations \cite{FWW2009,BS2009}, what is known as a distillation of non-locality. Introduction of the time ordered correlation classes allowed to confirm that the quantum correlations require multipartite information principles \cite{GWAN2011}. In what follows, it is aimed to find a new phenomena as well as applications connected with this subject.
 
In this paper the phenomenon of the non-locality emerging via wiring on 3-party boxes with binary inputs and binary outputs is studied. Definitions of the locality proposed in \cite{BBGP2011-definition} and \cite{GWAN2012-framework} differ in general. In \cite{GWAN2012-framework} a particular class of boxes closed under wiring is found. This class is called the time ordered bilocal one (TOBL). The property of closeness under wiring is crucial for the results presented here, thus we focus on the definition of multipartite non-locality from \cite{GWAN2012-framework}. Basing on the TOBL class, we introduce the counterparts of non-locality measures known for bipartite boxes - the {\it non-locality cost} and the {\it robustness of non-locality}. Subsequently, the analytical lower bounds on these measures in terms of the MWN are provided, namely, the maximum violation of the appropriate CHSH-like inequality \cite{CHSH} after application of the best wiring to some two of the three subsystems. This quantity, although may appear to be similar to the concept of N-copy distillable non-locality introduced in \cite{BCSS2011}, captures different properties of a box. The N-copy distillable non-locality quantifies how much non-locality can be obtained from the N-copies of a box using wiring transforming the N-boxes to a single box. The Maximal Wireable Non-locality is defined for a single copy of a multipartite box and wiring acting on some parties forming a group. 

We focus on the particular classes of boxes - the ones that admit the  particular model of a locality/non-locality, according to some grouping of the parties. Subsequently, we apply the Linear Programming to find the MWN for the considered classes.
An explicit example is the class of boxes which cannot be mapped to a non-local bipartite box by wiring applied to the two partitions (Bob and Charlie together, as well as Alice and Charlie together), while it can be mapped to a non-local bipartite box by some wiring applied to the third partition - Alice and Bob together.  If a quantum box with the analogous properties were found, it would serve as a resource for distributing non-locality in a local-like manner in analogy to distributing entanglement in separable manner \cite{CVDC-sep-ancilla} (see \cite{SKB-disord-bound,CMMPPP-disord-bound} for the quantitative description of this effect).

The original definition of locality by Svetlichny fails to fit into an operational framework of wiring, because the bipartite boxes which appear in the bilocal decomposition (\ref{eq-svd}) of a considered box are in a general signaling. The appearance of the signaling boxes is the reason for the non-locality emerging via wiring from such a box. In what follows, a subclass of boxes with bilocal decomposition (\ref{eq-svd}) is mostly considered. Namely, we focus on the particular cut: for example 3:12, when Alice (subsystem 1) and Bob (subsystem 2) are considered together and Charlie (subsystem 3) is at a distance. A box is {\it fully bilocal} in this cut if it can be expressed in a following way:
\ben
&&P(a_1,a_2,a_3|x_1,x_2,x_3) = \nonumber\\  &&\sum_{\nu} p_\nu  P_{\nu}(a_3|x_3)P_{\nu}(a_1,a_2|x_1,x_2),
\label{eq:bl}
\een
where $\sum_{\nu} p_\nu =1$. An upper bound on the MWN is given in terms of the weight of boxes signaling in the opposite direction to wiring which appear in fully bilocal decomposition. 

The paper is organized as follows. The section \ref{sec:intro} introduces the basic notions and useful parametrization of the tripartite non-signaling boxes, the CHSH values as well as wiring. The section \ref{sec:classes} begins with the comparison of known definitions of the local boxes, demonstrating explicitly that they are inequivalent and introduces classes of different {\it partially local} multipartite boxes, that is boxes which are fully bilocal in all the cuts as well as those that at least in one cut cannot be wired to a bipartite non-local box. The basic notions of the study are presented: the WN and the MWN. The \ref{subsec:sig-bound} provides an upper bound on the MWN for a particular box in terms of the weight of signaling boxes in its description according to fully bilocal decomposition (\ref{eq:bl}). The section \ref{subsec:known-facts} collects some known, useful facts about the non-locality cost for bipartite boxes with two binary inputs and two binary outputs. In \ref{subsec:3cost} we introduce 3-partite counterpart of non-locality cost, and show that linear function of the MWN places a lower bound on the 3-partite non-locality cost. Then an analogous result for 3-partite robustness of non-locality in section \ref{subsec:robustness} is demonstarted. Finally, the problem of finding the MWN for a given class of partially local boxes using the Linear Programming is studied. The particular boxes that allow to distribute the non-locality in a local-like manner (section \ref{subsec:3cost}) are found, as well as wiring with respect to its strength in creation of non-locality for different classes are classified.

\section{Tripartite non-signaling boxes and bipartite wiring}\label{sec:intro}

Any probability distribution belonging to the set of tripartite non-signaling correlations, with binary inputs  ($x_i$) and outputs ($a_i$) for each party, fulfils the following constraints:
\begin{eqnarray}
&& \forall\, a_1,a_2,a_3,x_1,x_2,x_3,\; P(a_1,a_2,a_3|x_1,x_2,x_3) \geq 0 \label{eq-positivity} \\
&& \forall x_1,x_2,x_3, \; \sum_{a_1,a_2,a_3}P(a_1,a_2,a_3|x_1,x_2,x_3) = 1 \; \label{eq-normalization}  \\
&&\forall a_2,a_3,x_2,x_3,x_1,x_1', \;\sum_{a_1}P(a_1,a_2,a_3|x_1,x_2,x_3)  \label{ns} \\ \nonumber &&= \sum_{a_1}P(a_1,a_2,a_3|x_1',x_2,x_3)   \nonumber, \\
&&\forall a_1,a_3,x_1,x_3,x_2,x_2', \;\sum_{a_2}P(a_1,a_2,a_3|x_1,x_2,x_3) \\ \nonumber &&= \sum_{a_2}P(a_1,a_2,a_3|x_1,x_2',x_3)   \nonumber, \\
&&\forall a_1,a_2,x_1,x_2,x_3,x_3', \;\sum_{a_3}P(a_1,a_2,a_3|x_1,x_2,x_3) \\ \nonumber &&= \sum_{a_3}P(a_1,a_2,a_3|x_1,x_2,x_3')   \nonumber.
\end{eqnarray}
 The set of tripartite boxes with the binary inputs and outputs, which satisfy these conditions, will be denoted as $NS_3$. The conditions presented above define a non-signaling polytope. It has been demonstrated \cite{PBS2011-3-polytop} that this polytope has 53 856 extremal points belonging to 46 different classes. All the deterministic extremal points form a single class, the remaining 45 classes consist of non-local extremal points. Due to the non-signaling and normalization constraints an arbitrary 3-partite box with binary inputs and outputs $P(a_1,a_2,a_3|x_1,x_2,x_3)$ can be written using the 26 parameters in the following way \cite{PBS2011-3-polytop}:
\begin{eqnarray}
\label{eq-bp}
&&P(a_1,a_2,a_3|x_1,x_2,x_3) = \\ \nonumber && \frac{1}{8} \left[ 1  + a_1 \left\langle A_{x_1}\right\rangle +a_2 \left\langle B_{x_2} \right\rangle  +  a_3\left\langle C_{x_3} \right\rangle   + a_1 a_2 \left\langle A_{x_1} B_{x_2} \right\rangle \right.   \\ \nonumber && + \left.  a_1 a_3 \left\langle A_{x_1} C_{x_3} \right\rangle + a_2 a_3 \left\langle B_{x_2} C_{x_3} \right\rangle +a_1 a_2 a_3 \left\langle A_{x_1} B_{x_2} C_{x_3} \right\rangle \right],
\end{eqnarray}
where e.g. $\left\langle A_{x_1} \right\rangle = P(a_1=1|x_1)-P(a_1=-1|x_1)$ is an expectation value of outcome for the input $x_1$. The notation where outputs ($\tilde{a},\tilde{b},\tilde{c}$) take values in $\{0, 1\}$ will be used. The relation between $\tilde{a},\tilde{b},\tilde{c}$ and $a,b,c$ is given by $a=(-1)^{\tilde{a}}$, $b=(-1)^{\tilde{b}}$, $c=(-1)^{\tilde{c}}$ \cite{PBS2011-3-polytop}. For the details of conversion of the expectation values to this notation see Appendix \ref{subsec:conversion}. From now on, for the sake of clarity, the $a$ would be written instead of $\tilde{a}$.

In what follows, the effect of {\it wiring} which maps tripartite boxes into bipartite ones will be studied. For this reason, the notion
of non-signaling {\it bipartite} boxes is also required. The latter boxes fulfil the following conditions:
\ben
&&\forall\, a_1,a_2,x_1,x_2\, P(a_1,a_2|x_1,x_2) \geq 0 \nonumber \\
&&\forall\, x_1,x_2\, \sum_{a_1,a_2} P(a_1,a_2|x_1,x_2) = 1 \nonumber \\
&&\forall\, a_2,x_2,x_1,x_1'\, \sum_{a_1} P(a_1,a_2|x_1,x_2) = \sum_{a_1} P(a_1,a_2|x_1',x_2) \nonumber\\
&&\forall\, a_1,x_1,x_2,x_2'\, \sum_{a_2} P(a_1,a_2|x_1,x_2) = \sum_{a_2} P(a_1,a_2|x_1,x_2') \nonumber \\
\een
The set of the non-signaling bipartite boxes with 2 binary inputs and 2 binary outputs will be denoted as $NS_2$.

\begin{figure}[h]
	\centering
		\includegraphics[width=70mm]{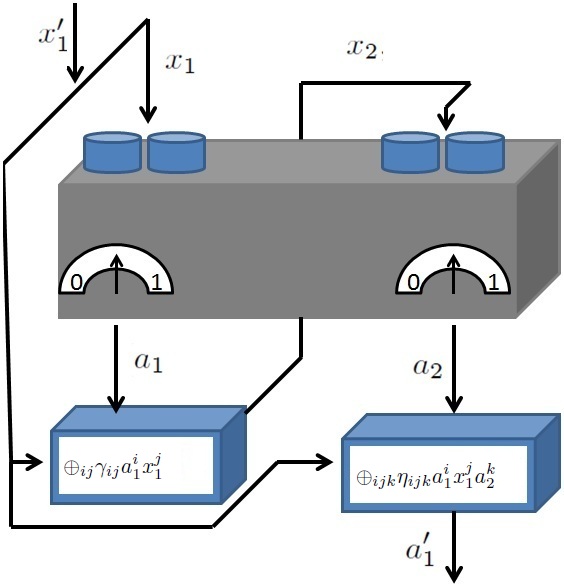}
		\caption{Depiction of wiring defined by vectors of binary coefficients $\gamma$ and $\eta$. An effective input bit is denoted as $x_1'$. This bit is equal to an input $x_1$ of the first party in a bipartition. An input $x_2$ of the party, which measures as the second one, is determined by the effective input bit $x_1'$ as well as an output of the first party $a_1$. An effective output bit depends on $x_1$', $a_2$ and an output of the second party $a_2$.}
	\label{fig1}
\end{figure}

Having the important sets of boxes introduced, we will focus on the wirings. It is possible to map a tripartite non-signaling box $P(a_1,a_2,a_3|x_1,x_2,x_3)$ into a bipartite one $P(a_1',a_3|x_1',x_3)$, having the bipartition as well as wiring set. 
The wiring presented in Figure \ref{fig1} would be considered. According to this wiring, the input of the first subsystem of the bipartite box - $x_1$ depends on $x_1'$ while the input of the second subsystem - $x_2$ can depend on $x_1'$ and the output of the first subsystem on - $a_1$. The effective output - $a_1'$ can depend on outputs $a_1, a_2$ and the effective input on $x_1'$.
 A particular parametrization of wiring will be used, in which an input of the first party in the bipartition will be \footnote{Other possibilities are either bit negation ($x_1=x_1' \oplus 1$) or choosing as an input constant bit (for example $x_1=0$)} $x_1=x_1'$. The second party chooses: 
\begin{equation}
x_2 = \oplus_{ij}\gamma_{ij}a_1^i x_1^j
\label{eq:delta}
\end{equation}
as an input, where $\oplus$ is an additional module 2, $\gamma_{ij}$ are binary constants and $i,\; j$ are also binary.
Similarly, the output of the box is defined as a polynomial of the form: 
\begin{equation}
a_1' = \oplus_{ijk}\eta_{ijk} a_1^i x_1^j a_2^k,
\label{eq:gamma}
\end{equation}
where $\eta_{ijk},\; i, \; j,\; k$ are binary. For a particular choice of $\gamma=(\gamma_{00},\ldots, \gamma_{11})$ and $\eta = (\eta_{000},\ldots,\eta_{111})$ wiring will be denoted as $W_{\gamma, \eta}$ or by specifying inputs and outputs $(x_2 = \oplus_{ij}\gamma_{ij}(a_1^i x_1^j), a_1' = \oplus_{ij}\eta_{ijk}( a_1^i x_1^j a_2^k) )$. To denote on which subsystem wiring is applied the following notation is used: $W^{X}_{\gamma,\eta}$ means that $W_{\gamma,\eta}$ is applied to systems $2$ and $3$, $W^{Y}_{\gamma,\eta}$ to systems $1$ and $3$ and finally $W^Z_{\gamma,\eta}$ to systems $1$ and $2$. It is also important to denote the order of measurements in particular wiring: for instance, on Fig. 1, the system 1 is measured prior to the system 2. Therefore by $W^{X_{\rightarrow}}_{\gamma,\eta}$ the second observer measures first and can send its results to the third observer. The parametrization (\ref{eq:delta}),(\ref{eq:gamma}) is valid for $W^{Z_{\rightarrow}}_{\gamma,\eta}$. In general, (\ref{eq:delta}),(\ref{eq:gamma}) should be modified accordingly to other choices of parties and/or ordering of measurements. In Section \ref{sec:case-study} it is argued that the number of the considered wiring can be restricted.

In order to verify if after the application of wiring effective probability distribution becomes non-local the value of one of the CHSH expressions \cite{Barret-Roberts} is calculated:

\ben
\nonumber
\beta_{rst}(P(a_1,a_2|x_1,x_2))= &&(-1)^{t}\left\langle 00\right\rangle + (-1)^{t+s} \left\langle 01\right\rangle  \\ \nonumber &&+ (-1)^{t+r} \left\langle 10\right\rangle  \\ &&+ (-1)^{t+s+r+1}\left\langle 11\right\rangle,
\label{eq:chsh}
\een

where $\left\langle ij\right\rangle = P(a_1 = a_2 | ij)- P(a_1 \neq a_2 | ij)$ and r, s, t take
values either 0 or 1. Correspondingly, the CHSH inequalities have a form: 
\ben
\label{eq:chshineq}
-2\leq \beta_{rst}(P(a_1,a_2|x_1,x_2)) \leq 2 
\een 
for binary $r,s,t$.
It is sufficient to consider the inequities (\ref{eq:chshineq}) equivalent to the CHSH, as the effective box after application of wiring to a tripartite box with all binary inputs and outputs is a box with two binary inputs and outputs. For more than binary inputs or outputs of a tripartite box it would be required to consider other Bell inequalities.

\section{Definitions of partially local multipartite boxes and the wire-emerging non-locality}
\label{sec:classes}
In this section, we present definitions of multipartite local boxes and justify the choice of the TOBL class.
As mentioned in the introduction, according to Svetlichny, no temporal order is imposed on bilocal terms in decomposition (\ref{eq-svd}), that is, signaling bipartite boxes can also appear in this decomposition. It has been recently noticed that signaling boxes may cause serious problems, since wired signaling probability distributions may lead to the grandfather-style paradoxes  \cite{BBGP2011-definition}. This fact has motivated the authors of \cite{BBGP2011-definition} to introduce the following definition of the partially mutipartite locality.

{\definition Correlations are $T_2$ local if $P(a_1,a_2,a_3|x_1,x_2,x_3)$ can be written in the form:
\ben
&&P(a_1,a_2,a_3|x_1,x_2,x_3) =\\ \nonumber  &&\sum_{\lambda} p_\lambda  P_{\lambda}(a_1|x_1)P^{2\rightarrow  3}_{\lambda}(a_2,a_3|x_2,x_3) + \\ \nonumber &&\sum_{\mu} p_\mu  P_{\mu}(a_2|x_2)P^{1\rightarrow  3}_{\mu}(a_1,a_3|x_1,x_3) + \\ \nonumber &&\sum_{\nu} p_\nu  P_{\nu}(a_3|x_3)P^{1\rightarrow  2}_{\nu}(a_1,a_2|x_1,x_2),
\label{eq:ts2}
\een
where $P_{\lambda,\mu,\nu}^{i\rightarrow j}(a_i,a_j|x_i,x_j)$ denotes probability distribution signaling at most in one direction, that is $\sum_{a_j} P_{\lambda,\mu,\nu}^{i\rightarrow  j}(a_i,a_j|x_i,x_j) = P_{\lambda,\mu,\nu}^{i\rightarrow  j}(a_i|x_i)$  and $ \sum_{a_i} P_{\lambda,\mu,\nu}^{j\rightarrow i}(a_i,a_j|x_i,x_j) = P^{j\rightarrow  i}_{\lambda,\mu,\nu}(a_j|x_j) $,  terms $P_{\lambda,\mu,\nu}^{i\rightarrow j}(a_i,a_j|x_i,x_j)$ can be replaced by $P_{\lambda,\mu,\nu}^{j\rightarrow i}(a_i,a_j|x_i,x_j)$ independently, $\sum_{\lambda} p_\lambda \geq 0, \sum_{\mu} p_\mu \geq 0, \sum_{\nu} p_\nu \geq 0$ and $\sum_{\lambda} p_\lambda + \sum_{\mu} p_\mu + \sum_{\nu} p_\nu =1$.} 

The above definition solves the problem of the time ordering. However, there is another definition that has been introduced from a different perspective. Namely, as it has been demonstrated in \cite{GWAN2012-framework}, if no time ordering of correlations is imposed, the "creation" of non-locality among N parties by means of local operations as well as classical communication is possible when N-1 parties collaborate. To avoid this type of misunderstanding the following definition has been proposed in \cite{GWAN2012-framework}. 

{\definition Correlations admit the TOBL model in cut $1:23$, when they can be written in a form:
\begin{eqnarray} \label{eq:def-classes}
&&P(a_1,a_2,a_3|x_1,x_2,x_3) = \nonumber\\  &&\sum_{\lambda} p_{\lambda} P^1_{\lambda}(a_1|x_1) P^{2\rightarrow3}_{\lambda}(a_2,a_3|x_2,x_3) = \\ \nonumber &&\sum_{\lambda} p_{\lambda} P^1_{\lambda}(a_1|x_1) P^{3\rightarrow2}_{\lambda}(a_2,a_3|x_2,x_3),
\end{eqnarray}
where $P_{\lambda}^{i\rightarrow j}(a_i,a_j|x_i,x_j)$ denotes probability distribution signaling at most in one direction. }

From the above definitions it can be seen that if a given box admits the TOBL model, it admits necessarily $T_2$ model. However, the converse is not true. One could, for instance, consider a box given in Table \ref{tab:ts2}. Following the procedure described in  \cite{BBGP2011-definition} one can verify that it belongs to $T_2$ class. Wiring $W_{\gamma, \eta}$ with $\gamma=(0,0,1,0)$ and $\eta=(0,1,0,0,0,0,0,0)$ applied to subsystems 1 and 2 with ($x_2=a_1,a_1'=a_2$) results in $P(a_2,a_3|x_1,x_3)$ for which $\beta_{000}(P(a_2,a_3|x_1,x_3))=\frac{7}{2}$ and it cannot belong to the TOBL.
\begin{table*}[t!]
\begin{ruledtabular}
\begin{tabular}{cccccccccccccc}
 \multicolumn{2}{c}{$\left\langle A_{x_1}\right\rangle$}&\multicolumn{2}{c}{$\left\langle B_{x_2}\right\rangle$}&\multicolumn{2}{c}{$\left\langle C_{x_3}\right\rangle$}&\multicolumn{4}{c}{$\left\langle A_{x_1} B_{x_2}\right\rangle$}&\multicolumn{4}{c}{$\left\langle A_{x_1} C_{x_3}\right\rangle$} \\
0&1&0&1&0&1&00&01&10&11&00&01&10&11\\
\hline
$-\frac{17783}{135743}$&$\frac{23193}{135743}$&$-\frac{195747}{542972}$&$\frac{212995}{542972}$&$\frac{35041}{542972}$&$\frac{19229}{542972}$&$-\frac{7097}{542972}$&$-\frac{10691}{542972}$&$-\frac{19295}{542972}$&$-\frac{8725}{542972}$&$\frac{291895}{542972}$&$-\frac{224737}{542972}$&$-\frac{252767}{542972}$&$-\frac{211635}{542972}$\\ 
\;&\;&\;&\;&\;&\;&\;&\;&\;&\;&\;&\; \\
&\multicolumn{4}{c}{$\left\langle B_{x_2} C_{x_3}\right\rangle$}&\multicolumn{8}{c}{$\left\langle A_{x_1} B_{x_2} C_{x_3} \right\rangle$}\\
00&01&10&11&000&001&010&011&100&101&110&111 \\ \hline
$\frac{25612}{135743}$&$\frac{51024}{135743}$&$-\frac{29459}{135743}$&$\frac{106063}{271486}$&$-\frac{110539}{135743}$&$\frac{65946}{135743}$&$\frac{115319}{135743}$&$-\frac{189937}{271486}$&$\frac{101225}{135743}$&$\frac{89089}{135743}$&$-\frac{108359}{135743}$&$-\frac{113289}{271486}$ \\
\end{tabular}
\end{ruledtabular}
\caption{An example of a box belonging to $T_2$ and not to the TOBL class. Definitions of the classes are presented in Sec. \ref{sec:classes}. For the details of the box parametrization see (\ref{eq-bp}) and (\ref{eq:binnot}).}  
\label{tab:ts2}
\end{table*}

In order to lower bound possible non-locality obtained by wirings in Sec. \ref{sec:nm} we introduce non-locality monotones. Their monotonicity is assured by the fact that certain sets involved in definitions of the monotones are closed under wirings. Due to this fact, we will use the TOBL model.
 In what follows we will focus on the three classes: the most general class of boxes with fully bilocal decomposition referred to as the S class, as it may include two-way signaling terms (\ref{eq:bl}), the TOBL ($T$) class and the NSBL class ($N$), defined that in (\ref{eq:def-classes}) only no signaling terms are allowed. It is known that \cite{GWAN2012-framework} 
\ben
\label{eq:classin}
NSBL \subset TOBL \subset S. 
\een

\subsection{Studied quantities and notation}

In what follows a box which in cut 1:23 and 2:13 belongs to the TOBL class and in cut 3:12 admits fully bilocal decomposition is considered (in general including signaling terms). This class of boxes will be denoted as TTS. Moreover, the article handles wiring acting on a subsystem 12 and also fixes the direction of wiring - from a subsystem 1 to 2.

\begin{figure}[h!]
	\centering
		\includegraphics[width=60mm]{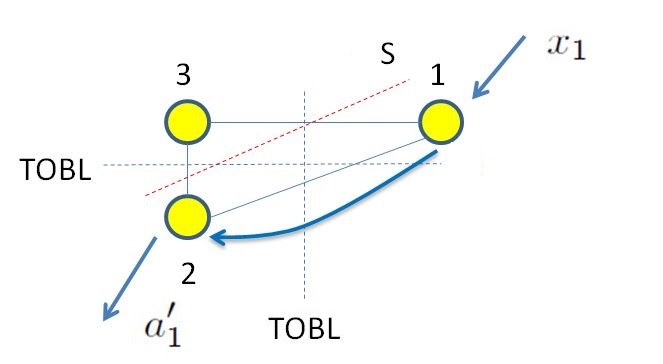}
		\caption{A TTS box for which quantities in the main text are defined. The blue dashed lines denote cuts in which the box admits the TOBL model (\ref{eq:def-classes}). The red line indicates fully bilocal decomposition (\ref{eq:bl})  in cut 3:12. Wiring acting on subsystems 1 and 2 in direction from a subsystem 1 to 2 is also depicted. }
\end{figure}

Using wiring two quantities could be defined: the WN and the MWN. The first one quantifies the violation of one of the CHSH inequalities (\ref{eq:chshineq}) that can be obtained using given wiring and the TTS class of correlations, whereas the second one gives the maximal violation of one of the CHSH inequalities that can be obtained using any wiring for the TTS class \cite{Barret-Roberts}.  
Formally, the WN for the given wiring $W_{\gamma,\eta}$ specified by some functions as in equation (\ref{eq:delta}) and (\ref{eq:gamma}) is defined as follows:  
\begin{eqnarray}
&&WN(W_{\gamma,\eta})= \left\{\begin{array}{l}\max_{P}
\beta_{000}(W_{\gamma,\eta}(P)):\\  \text{if}  \; \max_{P} \beta_{000}(W_{\gamma,\eta}(P))>2  \\0 \; : \; \text{otherwise}  \end{array} \right. \nonumber \\ \nonumber  && \text{subjected to} \; P \in TTS ,
\label{eq:wnw}
\end{eqnarray}
where P is a box from the TTS class. Only the violation of $\beta_{000}$ inequality is required to be considered, since the formula entails maximization over only the $2 \times 2  \times 2$ boxes. Indeed, if there is a box for which $|\beta_{000}| > 2$ then the same box after an appropriate local relabelling violates any other CHSH inequality (\ref{eq:chshineq}), as the latter equals the scalar product with a linear combination of the locally equivalent boxes \cite{Joshi-broadcasting}. In order to compute the WN, it would be tempting to follow \cite{Yang2011}, restricting the search to extremal vertices only, however the structure of different classes considered here is not yet known. The MWN for the the TTS class is: 
\begin{eqnarray}
MWN = \max_{\gamma, \eta} WN(W_{\gamma,\eta}).
\label{eq:mwnw}
\end{eqnarray}

Moreover, considering a single box P instead of the complete class of boxes, by means of the MWN for a given box, the maximal violation of one of the CHSH inequalities (\ref{eq:chshineq}) can be quantified. That can be obtained for a given cut of P using any wiring with mentioned order time of measurements. Formally, we write:
\ben
 MWN(P)= \max_{\gamma,\eta,r,s,t} \beta_{rst}(W_{\gamma,\eta}(P)).
\een
To simplify the notation, $\max_{\gamma,\eta} W_{\gamma,\eta}(.) \equiv \max_{W} W(.)$ and $\max_{r,s,t} \beta_{rst} (.) \equiv \max_{\beta} \beta(.)$ could be denoted.

\subsection{Upper bound on Maximal Wireable Non-locality from fully bilocal decomposition}\label{subsec:sig-bound}

In this section, the discussion centers on special boxes, namely the ones which are fully bilocal in cut 3:12 having form (\ref{eq:bl}) and do not belong to the TOBL class (\ref{eq:def-classes}) in this cut. It is also assumed that fully bilocal decomposition is explicitly known for the boxes under consideration, as it will be used to the upper bound MWN for these boxes.
 
Due to the fact that the non-zero non-locality after wiring involving systems 1 and 2 is caused by the signaling terms, it appears that the lower is the weight of such boxes in decomposition, the lower should be the MWN.

Here we follow this intuition and derive a bound on the MWN for a given box P described above in terms of the weight of boxes signaling in the opposite to wiring direction (in these considerations from subsystem 2 to subsystem 1) that appear in the fully bilocal decomposition.

Prior to demonstrating bound on the WN, we show that mere tracing out of the system cannot lead to a non-locality in case of the boxes with decomposition (\ref{eq:bl}). 

{\observation Consider a box $P(a_1,a_2,a_3|x_1,x_2,x_3)$ admitting decomposition (\ref{eq:bl}). For any $r,s,t \in\{0,1\}$, and for any value of $x_2 \in \{0,1 \}$ 
$$|\beta_{rst}(\sum_{a_2}P(a_1,a_2,a_3|x_1,x_2,x_3))| \leq 2.$$
\label{thm:trace}
}

{\it Proof}.

Because $P(a_1,a_2,a_3|x_1,x_2,x_3)$ is a legitimate box, the marginal distribution $P(a_1,a_3|x_1,x_3)=\sum_{a_2}P(a_1,a_2,a_3|x_1,x_2,x_3)$ is well defined. It is sufficient to demonstrate that the reduced box is local for fully bilocal decomposition. Therefore the attention can be focused on a particular input for the second party, for instance, the assumption that $x_2=0$. Then: 
\ben
\nonumber
&&P(a_1,a_3|x_1,x_3) = \sum_{a_2}P(a_1,a_2,a_3|x_1,0,x_3) = \\ \nonumber && \sum_{a_2, \nu} P_{\nu}(a_3|x_3)P_{\nu}(a_1,a_2|x_1,x_2)= \\  &&\sum_{\nu} P_{\nu}(a_3|x_3)P'_{\nu,x_2=0}(a_1|x_1).
\label{eq:t}
\een
The above equation demonstrates that the reduced box remains local. (Note that for $x_2=1$ terms $P'_{\nu,x_2=1}(a_1|x_1)$ may be different, however LHS of (\ref{eq:t}) will not change and the product form of RHS of (\ref{eq:t}) will be preserved). $\Box$

As a result, it is impossible to map a box admitting decomposition (\ref{eq:bl}) into a non-local one by a partial trace. However, it is not the case when one considers wiring as described in equations (\ref{eq:delta}), (\ref{eq:gamma}).

{\theorem 
Let $P(a_1,a_2,a_3|x_1,x_2,x_3)$ be a tripartite box with binary inputs and outputs admitting the fully bilocal decomposition: 
\ben
&&P(a_1,a_2,a_3|x_1,x_2,x_3) = \nonumber\\  &&\sum_{\nu} p_\nu P_{\nu}(a_3|x_3)P_{\nu}(a_1,a_2|x_1,x_2).
\label{eq:bl1}
\een
 The MWN of the box $P(a_1,a_2,a_3|x_1,x_2,x_3)$ satisfies the following bound:
\begin{equation}
MWN(P) = \max_{W,\beta} \beta (W (P)) \leq \inf_{p_{\nu^{s}}} 2 \sum_{\nu^s} p_{\nu^s} +2,
\label{eq:MWNP-bound}
\end{equation} 
where the maximum is taken over wiring $W$ with direction from subsystem 1 to 2 and $p_{\nu^{s}}$ are weights of bipartite boxes signaling opposite to direction, that is from a subsystem 2 to 1.
\label{thm:mwn-lbound}

}

{\it Proof}.

The wiring $W$ acting on subsystems 12 are again considered: 
\begin{eqnarray}
&&MWN(P) =\\ \nonumber &&\max_{W,\beta} \beta(W(P(a_1,a_2,a_3|x_1,x_2,x_3)) = \\ \nonumber 
&&\max_{W,\beta}  \beta(\sum_{\nu} p_\nu P_\nu (a_3|x_3)W(P_\nu(a_1,a_2|x_1,x_2)). \een
The MWN is independent from the particular decomposition of the form (\ref{eq:bl1}) which may not be unique.
For any decomposition, which is split into the two terms: the one signaling in direction of wiring and the one which does not, it leads to the following bound:
\ben
\label{eq:bd1}
&&\max_{W,\beta}  \\ \nonumber  &&\left[\sum_{\nu^{gs}} p_{\nu^{gs}} \beta(P_{\nu^{gs}} (a_3|x_3)W(P_{\nu^{gs}} (a_1,a_2|x_1,x_2))+\right. \\ \nonumber &&\left.
\sum_{\nu^{s}} p_{\nu^{s}} \beta(P_{\nu^{s}} (a_3|x_3)W(P_{\nu^{s}} (a_1,a_2|x_1,x_2))\right] \\ \nonumber && \leq 2\sum_{\nu^{gs}} p_{\nu^{gs}} + 4 \sum_{\nu^{s}} p_{\nu^{s}}  =  2 \sum_{\nu^{s}} p_{\nu^{s}} +2,
\end{eqnarray}
where $\sum_{\nu^{gs}}p_{\nu^{gs}}$ is the weight of boxes which are either non-signaling or singaling in the direction of the wiring. The fact that these boxes are mapped into local ones by wiring and $\sum_{\nu^s}p_{\nu^s} = 1 - \sum_{\nu^{gs}}p_{\nu^{gs}}$ was used. Hence, for each $\nu^{gs}$, $\beta$ on the bipartite box $P_{\nu^{gs}} (a_3|x_3)W(P_{\nu^{gs}} (a_1,a_2|x_1,x_2))$ emerging from wiring is bounded by 2.
Since the above inequality holds for any decomposition the one which leads to the tightest bound can be chosen, obtaining:
\ben
MWN(P) = \max_{W,\beta} \beta (W (P)) \leq \inf_{p_{\nu^{s}}} 2 \sum_{\nu^s} p_{\nu^s} +2,
\end{eqnarray}
where infimum is taken over decomposition (\ref{eq:bl1}) such that $\sum_{\nu^{s}}p_{\nu^{s}}$ is the weight of boxes which are singalizing in the direction opposite to wiring. $\Box$

Despite the fact that for signaling boxes the value of $\beta$ was replaced by its algebraic maximum, the bound (\ref{eq:MWNP-bound}) is tight.
   
In Table \ref{tab:pb} an example of the box for which the bound is tight is presented ($MWN(P)=\inf_{p_{\nu^{s}}} 2 \sum_{\nu^{s}} p_{\nu^{s}} +2 = 3$, for wiring acting on subsystems 1 and 2 ($x_2=a_1, a_1'=a_2$)).
\begin{table*}[t!]
\begin{ruledtabular}
\begin{tabular}{cccccccccccccccccccccccccccc}
\multirow{2}{*}{Class}& \; &\multicolumn{2}{c}{$\left\langle A_{x_1}\right\rangle$}&\multicolumn{2}{c}{$\left\langle B_{x_2}\right\rangle$}&\multicolumn{2}{c}{$\left\langle C_{x_3}\right\rangle$}&\multicolumn{4}{c}{$\left\langle A_{x_1} B_{x_2}\right\rangle$}&\multicolumn{4}{c}{$\left\langle A_{x_1} C_{x_3}\right\rangle$}&\multicolumn{4}{c}{$\left\langle B_{x_2} C_{x_3}\right\rangle$}&\multicolumn{8}{c}{$\left\langle A_{x_1} B_{x_2} C_{x_3} \right\rangle$}\\
&\;&0&1&0&1&0&1&00&01&10&11&00&01&10&11&00&01&10&11&000&001&010&011&100&101&110&111\\
\hline
&$1$&$0$&$0$&$-\frac{1}{20}$&$\frac{1}{20}$&$0$&$0$&$0$&$0$&$0$&$0$&$\frac{1}{20}$&$-\frac{1}{20}$&$\frac{1}{20}$&$-\frac{1}{20}$&$0$&$\frac{1}{2}$&$0$&$\frac{1}{2}$&$-\frac{1}{2}$&$\frac{1}{2}$&$\frac{1}{2}$&$-\frac{1}{2}$&$\frac{1}{2}$&$\frac{1}{2}$&$-\frac{1}{2}$&$-\frac{1}{2}$\\
\end{tabular}
\end{ruledtabular}
\caption{The exemplary box $P$ for which bound (\ref{eq:MWNP-bound}) on the $MWN(P)$ is tight. The bound yields $\inf_{p_{\nu^{s}}} 2 \sum_{\nu^{s}} p_{\nu^{s}} +2 = 3$. This value equals the $MWN(P)=3$ which is achieved for the wiring acting on subsystems 1 and 2 ($x_2=a_1, a_1'=a_2$). For the details of the box parametrization see (\ref{eq-bp}) and (\ref{eq:binnot}). }
\label{tab:pb}
\end{table*}

{\it Generalization of the notation. }
The quantities and the bound obtained in this section can be straightforwardly generalized as follows.
A tripartite box can have different kinds of correlations according to the different set of subsystems. 
A box $B_{XYZ}$ with $X,Y,Z \in \{S,T,N\}$ meaning that in partition where $23$ are together, it belongs to $X$ class, in partition where $13$ are together to $Y$ class and in partition where $12$ are together to $Z$ class. The set of all boxes with subindex $XYZ$ is called the $XYZ$-class. The class with some pattern of letters $T$ (or $N$) includes as a subset a class with another pattern of letters $T$ (or $N$), provided the latter can be obtained from the former by changing a {\it single} letter $T$ (or $N$) into $S$.
For instance:
\ben
TTT \subset TST \subset SST  \nonumber\\
NNN \subset NST \subset SST.   \label{eq:inclusions}
\een
By transitivity, sometimes even differing by two letters, it assures inclusion.
Other notation will be also required, namely $B_{1:23} \in TOBL$ means that the box $B$, when 2 and 3 are considered together,
belongs to the TOBL class. Then $B \in XYZ$ if $B_{1:23} \in X$, $B_{2:13} \in Y$ and $B_{3:12} \in Z$ is written.  
Wiring acting on different groups of a given box in arbitrary direction can be also considered. For instance, $W^{Z_{\rightarrow}}$ denotes wiring acting on subsystems 12, from a subsystem 1 to 2.

Taking into account the above notation one immediately generalizes all the introduced quantities. As an example, we give definition of the MWN for an arbitrary class and direction of wiring:   
\begin{eqnarray}
MWN_{XYZ} = \max_{\gamma, \eta,q_{\rightarrow}} WN_{XYZ}(W^{q_{\rightarrow}}_{\gamma,\eta}),
\end{eqnarray}
where $q$ denotes subsystems on which wring acts and $_{\rightarrow}$ takes into account direction of wiring.

\section{Non-locality monotones and wiring} \label{sec:nm}
Having considered the classes of partially local boxes, associated monotones, which measure multipartite non-locality with respect to a given class, can be defined. There will be two kinds of them: these that are counterpart of the (bipartite) cost of non-locality, and these that are counterpart of the bipartite (anti)robustness. Then it is demonstrated
that each of these multipartite monotones is lower bounded by the maximal violation of the appropriate CHSH-like inequality of the effective $2 \times 2$ box \cite{Barret-Roberts}. To derive 
the bound for multipartite case, the known results for $2\times 2$ bipartite boxes are first collected.

\subsection{Known properties of non-locality cost and twirlings for a $2\times 2$ case}\label{subsec:known-facts}
The non-locality cost in a $2\times 2$ case has the following definition:

{\definition \cite{EPR1992} The non-locality cost for a box P is defined as: 
\ben
\nonumber C(P) =&&\inf\{ \left.\right. \; p | P = pA + (1-p)L, \\ \nonumber &&\left.\right. {A \in  NS_2},\, L \in LR_{ns},\, p \in \left[0,1\right]\}, \\
\een
where P is a $2\times 2$ box, A denotes an arbitrary but no-signaling bipartite $2\times 2$ box and $LR_{ns}$ is the set of local non-signaling bipartite $2\times 2$ boxes.}

The non-locality cost
is monotonous under local operations. In particular, it is monotonous under twirling type operations $\tau_{rs}$.

{\definition \cite{Short,Nonsig_theories,Joshi-broadcasting}A twirling operation $\tau_{rs}$ is defined by flipping randomly 3 bits $\Delta_x,\Delta_y,\Delta_z$ and applying the following transformation to a $2\times 2$ box $P(a,b|x,y)$:

\ben
x &\rightarrow & x \oplus \Delta_x \nonumber \\
y &\rightarrow &y \oplus \Delta_y \nonumber \\
a &\rightarrow &a \oplus \Delta_y x \oplus \Delta_x \Delta_y \oplus \Delta_z \oplus s\Delta_y \nonumber \\
b &\rightarrow &b \oplus \Delta_x y \oplus \Delta_z \oplus r\Delta_x \nonumber. \\
\een
\label{def:twirling}}

It is known that the non-local vertices of the set $NS_2$ have the form \cite{Barret-Roberts}: 
\be B_{rst}(a, b|x, y) = \left\{\begin{array}{l} \frac{1}{2} \; \text{if} \; a\oplus b = xy\oplus r x \oplus s y \oplus t\\0 \; \text{else}  \end{array} \right. 
\ee
with $r,s,t = \{0,1\}$.

It is important, that after $\tau_{rs}$ any ($2\times 2$) non-signaling box becomes an {\it isotropic} box denoted as $P_{rs}^{\alpha}$ for some $\alpha \in [0,1]$, according to the following
parametrization:

\be P_{rst}^{\alpha}(a, b|x, y)=\alpha B_{rst}(a, b|x, y) + (1- \alpha)B_{rs\bar{t}}(a, b|x, y),\ee 

($\bar{.}$ denotes bit negation). Note that $B_{000}$ is a PR box and $B_{001}$ is an anti-PR box.
   The boxes $P_{rst}^{\alpha}$ are invariant under an appropriate twirling operation: $\tau_{rs}(P^{\alpha}_{rst})=P^{\alpha}_{rst}$. Adapting similar results as in \cite{KH-box-dist}, the following dependence occurs:
  
{\observation For a $2\times 2$ isotropic box $P_{rst}^{\alpha}$ with $\alpha \in [{1\over 2},1]$ and $t \in \{0,1\}$:
\be
C(P^{\alpha}_{rst})=\max\{0,4 \alpha-3\}
\ee
\label{obs:c-computed}
}
(for details of proof for $r=s=t=0$ see \cite{KH-box-dist}, for the other $r,s$ the proof is analogous). 

In what follows, another fact is also required, namely, that for any ($2\times 2$) box, $\beta_{rst}$ is invariant under $\tau_{rs}$ twirling operation \cite{context-measures}:
{\observation \label{obs:beta} \cite{KH-box-dist,Joshi-broadcasting} For any binary $r,s,t,r',s'$, a $2\times 2$ box P: 
\begin{equation}
\beta_{rst}(\tau_{r's'}(P)) = (8\alpha - 4)\delta_{r,r'}\delta_{s,s'}
\label{eq:beta-alpha}
\end{equation} 
and
\be
\beta_{rst}(P) = \beta_{rst}(\tau_{rs}(P))
\ee
for some $\alpha \in [0,1]$, where $\tau_{r's'}(P)=P^{\alpha}_{r's'}$ ($\tau_{rs}(P)=P^{\alpha}_{rs}$) denotes a box that is invariant under $\tau_{r's'}$ ($\tau_{rs}$) twirling operation, and $\delta$ is the Kronecker symbol.
}

Collecting the facts from the observations \ref{obs:beta} and \ref{obs:c-computed} as well as  using monotonicity of $C$ under local operations, the following fact is immediately obtained:

{\observation For a bipartite $2\times 2$ box $P$, and any $r,s\in\{0,1\}$, such that $\tau_{rs}(P) = P^{\alpha}_{rst}$, there is:
\begin{equation}
\begin{split}
C(P) \geq C(P^{\alpha}_{rst}) = &\frac{\beta_{rst}(P^{\alpha}_{rst})-2}{2} =\frac{\beta_{rst}(P)-2}{2}.
\end{split}
\label{eq:bandc}
\end{equation}
\label{obs:cost-beta}
}

\subsection{Non-locality cost for multipartite boxes and the lower bound}\label{subsec:3cost}

Considering in place of $LR_{ns}$ the class of partially local boxes $XYZ$ (that is such that at least one letter belongs to the set $\{TOBL,NSBL\}$), one obtains a measure of multipartite non-locality with respect to this class.
{\definition A non-locality cost for a box P with respect to the class XYZ is defined as:  
\ben
&&C_{XYZ}(P(a_1, a_2, a_3|x_1, x_2, x_3)) =\\ \nonumber &&\inf_{p} \left\{\; p | P(a_1, a_2, a_3|x_1, x_2, x_3) = pA + (1-p)L\right.,  \\  \nonumber && \left. A \in  NS_3, L \in XYZ ,\, p \in \left[0,1\right] \right\},
\een

where at least one of $X,Y,Z$ belongs to the set $\{NSBL,TOBL\}$ while the others are arbitrary in $\{NSBL,TOBL,S\}$.
} 

Following the dependences (\ref{eq:inclusions}), the relation is obtained:
\be
C_{XYZ} \geq C_{X'Y'Z'}
\ee
if $X'Y'Z'$ can be obtained from $XYZ$ by changing exactly one letter $T$ into $S$, in particular:
\ben
 C_{SST} \leq C_{TST} \leq C_{TTT} \nonumber \\
 C_{TSS} \leq C_{TTS} \leq C_{TTT}. 
\label{eq:C-inclusions}
\een
The $C_{XYZ}$ is non-increasing under linear operations, which preserves the set $XYZ$, namely, that transforms the set $XYZ$ into the set $XYZ$.

To lower bound of the multipartite non-locality cost, the attention will be centered on the non-locality cost with respect to the classes $TYZ$, that is, where in cut $1:23$ the box belongs to the TOBL or NSBL class, and set $Y$ and $Z$ are fixed arbitrarily:

\ben
&&C_{TYZ}(P(a_1, a_2, a_3|x_1, x_2, x_3)) =\\ \nonumber &&\inf_{p} \left\{ p | P(a_1, a_2, a_3|x_1, x_2, x_3) = pA + (1-p)L\right.,  \\  \nonumber && \left. A \in NS_3, L_{1:23} \in \text{TOBL}, L_{2:13} \in Y, L_{3:12}\in Z \right\},
\label{eq:def-nc}
\een
where $Y$ and $Z$ are arbitrary from the set $\{S,TOBL,NSBL\}$. Since these considerations will remain true for any choice of $Y$ and $Z$ the above
measure will be referred to as to $C_X$.

The following fact for the 3-party correlations can be observed:

{\lemma For any $2\times 2\times 2$ box $P$ $C_{X}(P)$ for $X \in \{NSBL,TOBL\}$ and $Y,Z \in \{NSBL,TOBL,S\}$, is lower bounded by the non-locality cost of a box 
emerging from $P$ under any wiring operation $W_{\gamma,\eta}$ applied to the systems $2$ and $3$ where the maximum over the directions of wiring W can be taken.
\label{lem:3cost}
}

{\it Proof. -} Let us fix $Y$ and $Z$ arbitrarily and wiring $W_{\gamma, \eta}$ on systems $23$ and its direction from the subsystem 2 to 3.
Let us assume, that $C(P(a_1, a_2, a_3| x_1, x_2, x_3))=p$, that is, $P(a_1, a_2, a_3| x_1, x_2, x_3) = pA + (1-p)L$. After applying the wiring, a $2\times 2$ box is obtained: $W_{\gamma, \eta}(P(a_1, a_2, a_3| x_1, x_2, x_3))=P(a_1, a_2'| x_1, x_2')\equiv P'$. By linearity of the wiring, $W_{\gamma, \eta}(P(a_1, a_2, a_3| x_1, x_2, x_3))=pW_{\gamma, \eta}(A) + (1-p)W_{\gamma, \eta}(L)$. Now, one recalls the fact that $L_{1:23} \in \{TOBL,NSBL\}$. It is known that the classes TOBL and NSBL are closed under wiring \cite{GWAN2012-framework}, thus the box $W_{\gamma,\eta}(L)$ is a local $2\times 2$ box. As a result,
the decomposition of $P'$ into  $W_{\gamma, \eta}(A)$ and $W_{\gamma, \eta}(L)$ is a valid decomposition into (possibly non-local) and local part, with the weight $p$ which can be then larger from $C(P')$. Hence, this is obtained:
\begin{equation}
C(P(a_1, a_2'| x_1, x_2')) \leq C(P(a_1, a_2, a_3| x_1, x_2, x_3)),
\label{costr}
\end{equation}
as desired.$\Box$

Having all the mentioned properties of the both 3- and 2- party non-locality, it can be seen that the non-locality cost for a $2\times 2\times 2$ box  is lower bounded by the linear function of the CHSH expression of a $2\times 2$ box resulting from the wiring:
{\theorem The non-locality cost for a $2\times 2\times 2$ box $P$ with $X\in\{TOBL,NSBL\}$, admits the following lower bound:
\begin{equation}
C_{X}(P) \geq \max_{W} C(W(P))  \geq  \max_{\beta, W}\frac{\beta(W(P))-2}{2} ,  
\label{eq:costlbound}
\end{equation} 
\label{thm:cost-lbound}
where the wiring acting on the subsystems 2 and 3 are considered and the maximum over direction of wiring is taken. 
}

{\it Proof}. - 

Let us fix $r,s \in \{0,1\}$, $\gamma,\eta$ and the direction of a wiring $W_{\gamma,\eta}$ arbitrarily. 
Denote $W_{\gamma,\eta}(P)$ as $P'$. 
First, lemma \ref{lem:3cost} is used to obtain: 
\ben
C_{X}(P) \geq C(P').
\een
By monotonicity of $C$ under $\tau_{rs}$ which is a {\it locality preserving operation} \cite{KH-box-dist}: 
\ben
C_{X}(P) \geq C(P') \geq C(\tau_{rs}(P')). 
\een
Now, there are two possibilities. First is that $C(\tau_{rs}(P')) =0$. Then the box $\tau_{rs}(P')$ is local, and hence $|\beta_{r's't'}(\tau_{rs}(P'))|\leq 2$ 
by definition for any $r's't'$. Then the second inequality in (\ref{eq:costlbound}) is satisfied.
 Second case is that $C(\tau_{rs}(P')) >0$.
 Then the box $\tau_{rs}(P')$ is not local which
in a $2\times 2$ case means that there exists a pair $r's'$ such that for all $t'$ there is $|\beta_{r's't'}(\tau_{rs}(P'))|>2$, since $\beta_{r's'0} = - \beta_{r's'1}$.
Now, the box $\tau_{rs}(P')$ is described as $\tau_{rs}(P') = P^{\alpha}_{rst}$ for $\alpha \in ({3\over 4},1]$ which fixes the value of $t\in\{0,1\}$.
Due to the observation \ref{obs:beta}: $r'=r$ and $s'=s$. Also $t' = t$ we obtain the following:
\be
\beta_{rst}(P^{\alpha}_{rst}) > 2.
\label{eq:good-beta}
\ee
Due to the observation \ref{obs:cost-beta}:
\be
C(P^{\alpha}_{rst}) \geq {\beta_{rst}(P') - 2 \over 2},
\ee
as desired. Due to the (\ref{eq:good-beta}) RHS of the above inequality is greater than zero.
From the above consideration, for any $r'',s'',t''\in\{0,1\}$ there is:
\be
{{\beta_{rst}(P') - 2}\over 2} \geq {{\beta_{r''s''t''}(P') - 2}\over 2}.  
\ee
Indeed, for $(r'',s'')\neq (r,s)$ RHS of the above equals $-1$, and for $(r'',s'') = (r,s)$, and $t''\neq t$, it is less than $-2$, while the LHS is positive by a
construction. This leads to:
\be
C_X(P) \geq C(W(P)) \geq \max_\beta {{\beta(W(P)) - 2}\over 2}.
\ee
Since $W_{\gamma,\eta}$ was arbitrary, maximising over the wiring, the desired chain of inequalities is obtained. $\Box$

If the class of partially local boxes has more than one cut which admits the TOBL or the NSBL model, then the above theorem can be applied to these cuts, and obtain independent lower bounds. Taking 
supremum over the cuts yields a superior lower bound, hence an immediate corollary is obtained:

{\corollary Let $Q \subset \{X,Y,Z\}$ such that for $q \in Q$ there is $q \in \{TOBL,NSBL\}$. Then, for any tripartite box $P$, there is:
\be
C_{XYZ}(P) \geq \max_{q \in Q} \max_{\beta, W^{q}} {\beta(W^{q}(P)) - 2\over 2},
\ee
\label{cor:cost-bound}
where the maximum over the direction of wiring is taken.
}

\subsection{Multipartite robustness of non-locality and the lower bound}\label{subsec:robustness}
In analogy to non-locality cost the so called {\it robustness} $R$ is studied which is a multipartite counterpart of the measure given by $R \equiv 1-\bar{R}$ \cite{Joshi-broadcasting} which is defined as follows:

{\definition For a bipartite $2\times 2$ box $P \in NS_2$, its robustness of non-locality  is defined as:
\begin{equation}
R(P) = \inf_{A \in NS_2} \left\{p \left|\right. pA+(1-p)P \in LR_{ns}  ,\, p \in \left[0,1\right] \right\},
\label{eq:def-rob}
\end{equation}
where A is an arbitrary bipartite non-signaling $2 \times 2$ box and $LR_{ns}$ denotes the set of non-signaling local-realistic bipartite $2 \times 2$ boxes.
}

The multipartite robustness of non-locality for 3 parties is defined with respect to a class of local boxes $XYZ$ (that is such that at least one letter belongs to the set $\{TOBL,NSBL\}$):
{\definition
For a tripartite $2\times 2\times 2$ box $P \in NS_3$, its robustness of non-locality with respect to a class of local boxes $XYZ$, where at least
one of $X,Y,Z$ belongs to $\{NSBL,TOBL\}$ and the others are arbitrary in $\{NSBL,TOBL,S\}$ reads:
\ben
R_{XYZ}(P) = &&\inf_{A\in NS_3} \left\{p \left|\right. pA+(1-p)P \in XYZ\right. \\ \nonumber &&\left. p \in \left[0,1\right] \right\},
\een
where the infimum is taken over the arbitrary non-signaling $2\times 2\times 2$ boxes.
}

Similarly as for non-locality cost (\ref{eq:C-inclusions}), the following dependencies occur:
\ben
R_{TTT} \geq R_{TST} \geq R_{SST}  \nonumber \\
R_{TTT} \geq R_{TTS} \geq R_{TSS}.  
\label{eq:R-inclusions}
\een

Since considerations concerning bound on this measure are analogous to that for cost of non-locality, just the results are here presented. For the sake of completeness, the proofs are presented in
Appendix. In analogy to lemma \ref{lem:3cost} it is demonstarted that multipartite robustness does not increase under wiring:

{\lemma For any $2\times 2\times 2$ box $P$, its robustness $R_{XYZ}(P)$ for $X \in \{NSBL,TOBL\}$ and $Y,Z \in \{NSBL,TOBL,S\}$, is lower bounded by the robustness of non-locality of a box 
emerging from $P$ under wiring operation applied to systems $2$ and $3$.
\label{lem:robustness}}

Realizing this, an analogue of theorem \ref{thm:cost-lbound} for multipartite robustness can be stated:

{\theorem The robustness of non-locality for a $2\times 2\times 2$ box $P$ with $X\in\{TOBL,NSBL\}$, admits the following lower bound:
\ben
\label{eq:robustness}
&&R_{XYZ}(P) \geq \max_{W} R(W(P))\geq \\ \nonumber
&&\max_{\beta,W} \frac{\beta (W(P))-2}{\beta(W(P))+4},   
\een where wiring W acting on subsystems 2 and 3 is considered and the maximum is taken over wiring direction. 

\label{thm:robustness}
}

Finally, if the class of boxes is closed under wiring with respect to more than one cut, the bounds over the MWN in these cuts can be maximized:

{\corollary Let $Q \subset \{X,Y,Z\}$ such that for $q \in Q$ there is $q \in \{TOBL,NSBL\}$. Then, for any tripartite $2 \times 2 \times 2$ box $P$, there is:
\be
R_{XYZ}(P) \geq \max_{q \in Q} \max_{\beta, W^{q}} {\beta(W^{q}(P)) - 2\over \beta(W^{q}(P)) + 4},
\ee
where the maximum over the direction of wiring is taken.
}

\section{The MWN - the case study via the Linear Programming}\label{sec:case-study}
In this section the MWN is studied by the Linear Programming. Different classes of tripartite boxes are considered: NNS, NTS, TTS, NSS and TSS class. Prior to presenting results it is sufficient to restrict the considerations to wiring of a simple form.

{\observation \label{obs:res}
Let us consider a wiring $W_{\gamma,\eta}$. For the WN the following relation holds $WN(W_{\gamma,\eta})=WN(W_{\bar{\gamma},\eta})$ where $W_{\bar{\gamma}, \eta}$ denotes wiring of a simple form, namely, $(x_2 =a_1, a_1' = \oplus_{ij}\eta_{ijk}( a_1^i x_1^j a_2^k) )$. Moreover, $MWN = \max_{\eta}(W_{\bar{\gamma}, \eta})$. 
 }

The idea of the proof of the above observation (see Appendix sec. \ref{subsec:proofs}) is based on the fact that the action of any wiring $W_{\gamma, \eta}$ on a box $P$ can be implemented by the wiring of the simple form $W_{\bar{\gamma},\eta}$ on a box $P'$ that can be obtained from $P$ by the local operations. Hence, since the $WN$ entiles maximization over all boxes from the same class, to which $P$ and $P'$ belongs, it is sufficient to consider the wiring of the simple form.

The values of the $MWN$ and WN for different classes are summarized in the Table \ref{tab:classes}. Only the presented classes matter, as far as the MWN of a class is concerned, since classes with the same number of letters T (or N) yield the same MWN. Moreover, due to the relations (\ref{eq:classin}) the MWN for SSS class can be obtained from a Table \ref{tab:classes} (because $TSS \subset SSS$ and the MWN for TSS class yields maximal possible value for bipartite boxes with binary inputs and outputs). We found that as far as wirings are concerned, there is no difference between the NNS and NTS. The same holds for the TSS class.

\begin{table}[h!]
\begin{tabular}{|c|c|c|}
\hline
Class&$MWN$&$WN(W_{\bar{\gamma}, \eta})$ \\
\hline
NTS, NNS&3&$2\frac{4}{5}$ \\
\hline
TTS&3&$2\frac{12}{13}$\\
\hline
NSS, TSS&4&3\\
\hline
\end{tabular}
\caption{The MWN for different correlations classes (N - NSBL, T -TOBL, S- boxes with the fully bilocal decomposition which may entile the two way signaling boxes). The values of the WN are also presented.}
\label{tab:classes}
\end{table}

The wiring leading to the non-zero WN for classes are presented in Tables \ref{tab:nns}, \ref{tab:tts}, \ref{tab:nss}, \ref{tab:tss}. The full list of wiring can be obtained from these Tables by performing the local relabeling of $a_2$ ($a_2\rightarrow a_2+1$ and $a_2 \rightarrow a_2+a_1$). 
For a given class, in principle there could be as many WNs as non-trivial wiring, however, as it can be seen in Tables \ref{tab:nns}, \ref{tab:tts}, \ref{tab:nss}, \ref{tab:tss} there are only two of them. Having a given value of the WN, it could be that depending on wiring, a different box to attain it is required. Interestingly, we have found a box which we call a representative for this WN, as for any wiring its WN can be obtained on some local relabeling of this box.
For instance, there are two representative boxes for the NSS class for the two values of the WN: 3 and 14/5 (see Table \ref{tab:rtts} and  \ref{tab:btts} for the analogous results for the TTS class). For the representative boxes the upper bound on the MWN of the Theorem \ref{thm:mwn-lbound} is computed, as well as lower bounds on non-locality cost and robustness. These results are presented in Tables \ref{tab:rtts} and \ref{tab:rnns}, for the TTS and NNS class respectively. For the representative box 1 in the case of the TTS and the NNS correlations the upper bound on the MWN is tight. For the other classes of correlations we have not succeeded in finding representative box. 

\begin{table*}[t!]
\begin{ruledtabular}
\begin{tabular}{c|ccccccccccccccccccccccccccc}
\multirow{2}{*}{Box}& \; &\multicolumn{2}{c}{$\left\langle A_{x_1}\right\rangle$}&\multicolumn{2}{c}{$\left\langle B_{x_2}\right\rangle$}&\multicolumn{2}{c}{$\left\langle C_{x_3}\right\rangle$}&\multicolumn{4}{c}{$\left\langle A_{x_1} B_{x_2}\right\rangle$}&\multicolumn{4}{c}{$\left\langle A_{x_1} C_{x_3}\right\rangle$}&\multicolumn{4}{c}{$\left\langle B_{x_2} C_{x_3}\right\rangle$}&\multicolumn{8}{c}{$\left\langle A_{x_1} B_{x_2} C_{x_3} \right\rangle$}\\
&\;&0&1&0&1&0&1&00&01&10&11&00&01&10&11&00&01&10&11&000&001&010&011&100&101&110&111\\
\hline
&$1$&$0$&$0$&$0$&$0$&$0$&$0$&$0$&$0$&$0$&$0$&$0$&$0$&$0$&$0$&$0$&$\frac{1}{2}$&$0$&$\frac{1}{2}$&$-\frac{1}{2}$&$\frac{1}{2}$&$\frac{1}{2}$&$-\frac{1}{2}$&$\frac{1}{2}$&$\frac{1}{2}$&$-\frac{1}{2}$&$-\frac{1}{2}$\\
&$2$&$\frac{1}{13}$&$\frac{1}{13}$&$-\frac{3}{13}$&$\frac{5}{13}$&$\frac{1}{13}$&$-\frac{1}{13}$&$\frac{1}{13}$&$\frac{1}{13}$&$\frac{1}{13}$&$\frac{5}{13}$&$\frac{9}{13}$&$-\frac{5}{13}$&$\frac{1}{13}$&$-\frac{1}{13}$&$\frac{5}{13}$&$\frac{7}{13}$&$\frac{5}{13}$&$\frac{7}{13}$&$-\frac{3}{13}$&$\frac{3}{13}$&$\frac{5}{13}$&$-\frac{5}{13}$&$\frac{5}{13}$&$\frac{7}{13}$&$-\frac{7}{13}$&$-\frac{5}{13}$\\
\end{tabular}
\end{ruledtabular}
\caption{The representative boxes from the TTS class associated to the groups of wiring.}
\label{tab:rtts}

\begin{ruledtabular}
\begin{tabular}{c|c|c|c|c|c}
Box&Representative wiring&$WN(W_{\gamma, \eta})$&Upper bound on $MWN$&Lower bound on non-locality cost&Lower bound on robustness\\
\hline
1&$a_1,a_2$&$3$&3&$C\geq\frac{1}{2}$&$R\geq\frac{1}{7}$\\
2&$a_1,a_2+a_1 a_2 x_1$&$\frac{38}{13}$&$\frac{50}{13}$&$C\geq\frac{6}{13}$&$R\geq\frac{2}{15}$
\end{tabular}
\end{ruledtabular}
\caption{Wiring that together with the boxes from Table \ref{tab:rtts} attain the maximum of the WN (local relabeling of boxes is not required). In subsequent columns the value of the WN, upper bound on the MWN for boxes from Table \ref{tab:rtts}, as well as Lower bound on non-locality cost and robustness are presented.}
\label{tab:btts}
\end{table*} 

\begin{table*}[t!]
\begin{ruledtabular}
\begin{tabular}{c|ccccccccccccccccccccccccccc}
\multirow{2}{*}{Box}& \; &\multicolumn{2}{c}{$\left\langle A_{x_1}\right\rangle$}&\multicolumn{2}{c}{$\left\langle B_{x_2}\right\rangle$}&\multicolumn{2}{c}{$\left\langle C_{x_3}\right\rangle$}&\multicolumn{4}{c}{$\left\langle A_{x_1} B_{x_2}\right\rangle$}&\multicolumn{4}{c}{$\left\langle A_{x_1} C_{x_3}\right\rangle$}&\multicolumn{4}{c}{$\left\langle B_{x_2} C_{x_3}\right\rangle$}&\multicolumn{8}{c}{$\left\langle A_{x_1} B_{x_2} C_{x_3} \right\rangle$}\\
&\;&0&1&0&1&0&1&00&01&10&11&00&01&10&11&00&01&10&11&000&001&010&011&100&101&110&111\\
\hline
&$1$&$0$&$0$&$-\frac{1}{20}$&$\frac{1}{20}$&$0$&$0$&$0$&$0$&$0$&$0$&$\frac{1}{20}$&$-\frac{1}{20}$&$\frac{1}{20}$&$-\frac{1}{20}$&$0$&$\frac{1}{2}$&$0$&$\frac{1}{2}$&$-\frac{1}{2}$&$\frac{1}{2}$&$\frac{1}{2}$&$-\frac{1}{2}$&$\frac{1}{2}$&$\frac{1}{2}$&$-\frac{1}{2}$&$-\frac{1}{2}$\\
&$2$&$-\frac{2}{5}$&$\frac{1}{5}$&$-\frac{3}{5}$&$\frac{1}{5}$&$\frac{1}{5}$&$-\frac{1}{5}$&$0$&$0$&$\frac{1}{5}$&$\frac{1}{5}$&$0$&$-\frac{2}{5}$&$\frac{1}{5}$&$-\frac{1}{5}$&$\frac{1}{5}$&$\frac{3}{5}$&$\frac{1}{5}$&$\frac{3}{5}$&$-\frac{2}{5}$&$0$&$\frac{2}{5}$&$-\frac{4}{5}$&$\frac{1}{5}$&$\frac{3}{5}$&$-\frac{3}{5}$&$-\frac{1}{5}$\\
\end{tabular}
\end{ruledtabular}
\caption{The representative boxes from the NNS class associated to the groups of wiring. }
\label{tab:rnns}
\begin{ruledtabular}
\begin{tabular}{c|c|c|c|c|c}
Box&Representative wiring&$WN(W_{\gamma,\eta})$&Upper bound on $MWN$&Lower bound on non-locality cost&Lower bound on robustness\\
\hline
1&$a_1,a_2$&$3$&3&$C\geq\frac{1}{2}$&$R\geq\frac{1}{7}$\\

2&$a_1,a_1+a_1 a_2 x$&$\frac{14}{5}$&$\frac{18}{5}$&$C\geq\frac{2}{5}$&$R\geq\frac{1}{17}$

\end{tabular}
\end{ruledtabular}
\caption{Wiring that together with boxes from Table \ref{tab:rnns} attains the maximum of the WN (local relabeling of boxes is not required). In subsequent columns the value of WN, upper bound on the MWN for boxes from Table \ref{tab:rnns}, as well as lower bound on non-locality cost and robustness are presented.}
\label{tab:bnns}

\end{table*} 

In some cases, using wiring, it is possible to determine to which class of correlations a given box cannot belong to. If $\beta_{rst}(W_{\gamma, \eta}(P))$ is higher than the value of $WN_{XYZ}(W_{\bar{\gamma}, \eta})$ for some XYZ, it implies that $P \not \in XYZ$. For instance, if after any wiring from a Table \ref{tab:nns} with the $WN =\frac{14}{5}$ value of any CHSH expression is higher than this WN, the box cannot belong to the NNS and NTS class.

\begin{figure}[h!]
	\centering
		\includegraphics[width=100mm]{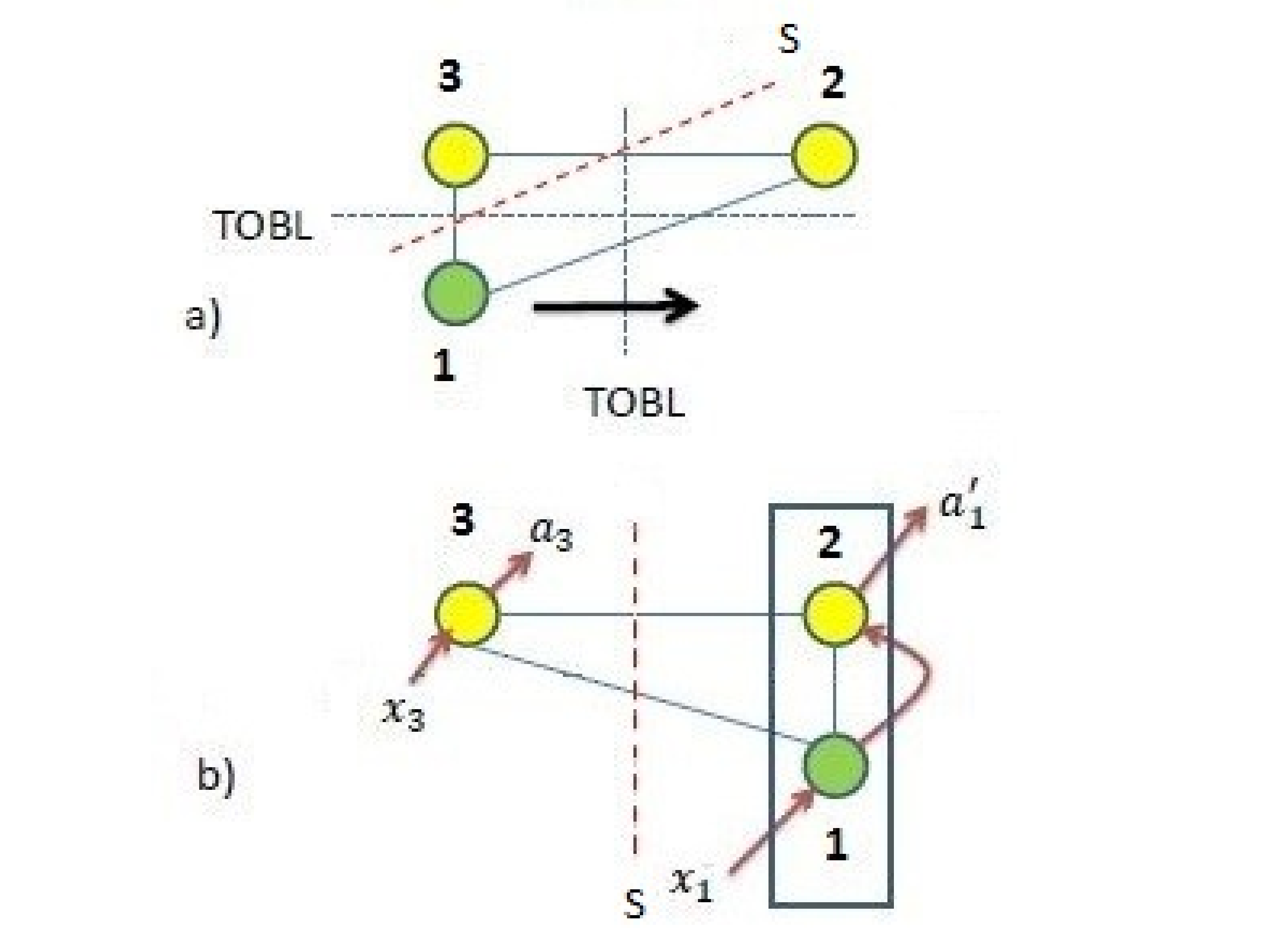}
		\caption{A box-analogue of distribution of entanglement by separable ancilla. a) A box from the TTS class, such that no non-locality can be created by wiring in cuts 1:23 and 2:13. In cut 3:12 this box admits fully bilocal decomposition. b) When the system 1 is transferred to 2, after wiring on systems 12, the initial tripartite box $P(a_1,a_2,a_3|x_1,x_2,x_3)$ becomes an effective bipartite, {\it non-local} box $P(a_1',a_3|x_1,x_3)$.}
	\label{C-box}
\end{figure}

{\it Distributing non-locality in a local-like manner.-}
In \cite{CVDC-sep-ancilla} it is shown that one can distribute entanglement "without entanglement": using ancillary state, that in each step is separable with the rest of the system. Correlations belonging to the NNS, NTS and TTS class are such that in cut 1:23, and 2:13 they cannot be wired to a non-local box, while in cut 3:12, after suitable wiring, the WN is non-zero. Therefore using these boxes and the appropriate wiring it is possible to distribute non-locality in a local-like manner. For instance, it could be that initially system 2 is possessed by one party (Alice) and systems 13 are possessed by the other one (Bob). Then no non-locality between Alice and Bob can be created by the wiring applied to the Bob's devices. The situation transforms when the system 1 is transferred from Bob to Alice (see Figure \ref{C-box}). Then, after applying wiring to subsystem 12, the effective box shared by Alice and Bob will become non-local. 

It is often the case in the Quantum Information Theory that new resources and (more or less) real life scenarios, become related. The Quantum Key Distribution is the most profound example of such an approach \cite{BB1984}. Suppose that in the NNS/NTS/TTS there is a box which after wiring is useful for the so called Device Independent QKD. Then the contrived, whereas still possible cryptographic scenario can be introduced. Consider a situation in which an agent Alice would like to communicate with an agent Bob in a secure way. One of the possible solutions would be to equip the both agents with the secure devices. However, due to the character of her activity, Alice may be caught and her device may be at some point in the hands of enemies.  In order to prevent enemies from using her device, one more element of the system would be desirable. This element is kept in a secure location C and enables security. Such a tripartite system can be built from the box, which was mentioned above, belonging to the NNS/NTS/TTS class. Let us focus on a box belonging to the TTS class presented in Figure \ref{C-box}.  One equips agent Alice with a pocket device consisting of a subsystem 2 of the considered box, another part of the device with subsystem 1 is kept in the secure location known to Alice, and agent Bob is in possession of a pocket device with a subsystem 3. When Alice has an access to the two specific subsystems of the complete box (a subsystem 2 in her pocket and a subsystem 1 in the secure location C), she can perform appropriate wiring and then the effective box shared by Alice and Bob will become non-local. Otherwise, due to the observation \ref{thm:trace}, the box shared by Alice and Bob (consisting of subsystems 2 and 3) is local and therefore it cannot be used to perform any cryptographic task. From the security point of view in order to set up such a system against quantum adversary one can also use the quantum states from \cite{CVDC-sep-ancilla}, provided that secure key can be extracted from them.

\section{Conclusions}

The phenomenon of non-locality emerging from the application of wiring involving 2 parties, to a 3-partite boxes with binary inputs and outputs have been studied quantitatively. In particular, the natural counterparts of the known bipartite non-locality measures, such as cost of non-locality and its robustness have been introduced, placing a lower bound on these measures in the terms of explicit functions of maximal violation of the CHSH inequality after wiring. 

Presented approach can be generalized to the case of a tripartite non-signaling box with a larger number of inputs and/or outputs. However, then the Bell expressions other than the CHSH must be considered. It is also straightforward to generalize these definitions to the multipartite case with $m\geq 4$ parties, however the bounds should then involve the violation of some multipartite Bell inequality for more than 2 parties.

The class of partially local boxes can be defined in a more general way, that is, as boxes admitting in bipartition the fully bilocal decomposition.  

 We have shown, that maximal attainable non-locality via wiring is upper bounded by the weight of boxes signaling in opposite direction to the wiring in fully bilocal decomposition of a box. It would be also interesting to place some lower bounds based on this description. Finally, we have studied the MWN using the Linear Programming. In particular, we have identified the boxes which fall into an interesting class enabling the distribution of non-locality in a local manner. The boxes which maximize the value of the WN in each of the considered classes are manifestly non-quantum (reaching $3 > 2\sqrt{2}$ of violation of the CHSH inequality). It would be interesting to find their quantum-realizable versions like the one demonstrated in \cite{GWAN2012-framework}. We have also classified different wiring proving that some of them are equivalent as far as the increase of the non-locality under their application is concerned. 
These findings shed some light on the phenomenon of non-locality emerging from processing of the multipartite non-locality via wiring.

\textbf{Acknowledgments.-} The authors acknowledge the discussions with Pawe\l{} Horodecki and Antonio Acin, and thank to Micha\l{} Horodecki for the helpful comments.  The calculations were conducted at the Academic Computer Center in Gda\'nsk. K.H. and J.T. acknowledge the grant of the Polish Ministry of Science and the Higher Education Grant no. IdP2011 000361. J.T. is additionaly supported by the ERC AdG grant QOLAPS.

\section{Appendix}
\subsection{The conversion of expectation values}\label{subsec:conversion}
The conversion of expectation values between notation in which $a_1,a_2,a_3, \in \{-1,1 \}$ and $\tilde{a}_1,\tilde{a}_2,\tilde{a}_3 \in \{ 0,1 \}$ is given by \cite{PBS2011-3-polytop}:
\ben
\label{eq:binnot}
&&\left\langle A_{x_1}\right\rangle = 1-2\left\langle \tilde{A}_{x_1}\right\rangle \\ \nonumber
&&\left\langle A_{x_1} B_{x_2}\right\rangle =  1-2\left\langle \tilde{A}_{x_1}+\tilde{B}_{x_2}\right\rangle \\ \nonumber
&&\left\langle A_{x_1} B_{x_2} C_{x_3}\right\rangle =  1-2\left\langle \tilde{A}_{x_1}+\tilde{B}_{x_2}+\tilde{C}_{x_3}\right\rangle, \\ \nonumber 
&&\text{where} \\ \nonumber
&&\left\langle \tilde{A}_{x_1}\right\rangle = \sum_{\tilde{a}_1} P(\tilde{a}_1|x_1)\tilde{a}_1 \\ \nonumber
&&\left\langle \tilde{A}_{x_1}+ \tilde{B}_{x_2}\right\rangle = \sum_{\tilde{a}_1\; \tilde{a}_2 } P(\tilde{a}_1 \tilde{a}_2|x_1 x_2)\tilde{a}_1\tilde{a}_2 \\ \nonumber
&&\left\langle \tilde{A}_{x_1}+ \tilde{B}_{x_2}+\tilde{C}_{x_3}\right\rangle = \sum_{\tilde{a}_1\; \tilde{a}_2 \; \tilde{a}_3 } P(\tilde{a}_1 \tilde{a}_2 \tilde{a}_3|x_1 x_2 x_3)\tilde{a}_1\tilde{a}_2 \tilde{a}_3.
\een

\subsection{Proofs and examples}\label{subsec:proofs}
In this section the details of the proof of lemma \ref{lem:robustness} and theorem \ref{thm:robustness} are demonstrated. We initiate with some useful facts about robustness for the bipartite case.
 In particular, it was shown in \cite{Joshi-broadcasting} that for isotropic boxes $P^{\alpha}_{rs}(a_1, a_2| x_1, x_2) = \alpha B_{rst}(a_1, a_2| x_1, x_2) + (1- \alpha)B_{rs\bar{t}}(a_1, a_2| x_1, x_2)$ there is: 
\be
\bar{R}(A)=\frac{3}{4 \alpha} 
\ee  
and for A such that $\beta_{rst}(A)\geq 2$ there is:
\be
\bar{R}(A)=\bar{R}(\tau_{rs}(A)).
\label{eq:r-inv-t}
\ee
From this fact, one obtains that for $\alpha > 3/4$:
\be
R(P_{rs}^{\alpha})=\frac{4 \alpha -3}{4\alpha} = \frac{\beta_{rst}(P_{rs}^{\alpha})-2}{\beta_{rst}(P_{rs}^{\alpha})+4}
\label{eq:r-beta}
\ee
for any binary $r,s,t$.

Having collected the known facts for robustness in a $2\times 2$ case lemma \ref{lem:robustness} can be proved which states that robustness of a $2 \times 2 \times 2$ box is monotonous under wiring on two subsystems. It is demonstrated for the class $X$, as for other classes the proof is analogous.

{\it Proof of lemma \ref{lem:robustness}} 

Let us fix $Y$ and $Z$ arbitrarily and wiring $W_{\gamma, \eta}$ on systems $2$ and $3$ with arbitrary direction. It could be assumed that $R(W_{\gamma,\eta}(P))=\tilde{p}$ 
and $R(P)=p$. Then, there exists a box $L \in \{NSBL,TOBL\}$ with respect to 1:23 cut, such that 
$p X+ (1-p) P = L$. By linearity of wiring the following is obtained:
\be p W_{\gamma, \eta}(X)+ (1-p) W_{\gamma, \eta}(P) = W_{\gamma, \eta}(L).
\label{eq:dec}\ee Now, by the fact that classes $N$ and $T$ yield local boxes under wiring on $2$nd and $3$rd subsystems, it is obtained that $W_{\gamma, \eta}(L)=L'$ is a $2\times 2$ local box. Hence, the decomposition (\ref{eq:dec}) is valid decomposition of a local box $L'$ into $W_{\gamma,\eta}(P)$ and some other box which
confirms $p\geq \tilde{p}$ as expected.$\Box$

As a result the Robustness for a $2\times 2\times 2$ box is lower bounded by the linear function of the CHSH expression of a $2\times 2$ box resulting from wiring
which is stated in the theorem \ref{thm:robustness}. The proof of this theorem is presented below:

{\it Proof of the theorem \ref{thm:robustness} }. Let us fix $\gamma,\eta$ arbitrarily. Then, by lemma \ref{lem:robustness}, $R_{XYZ}(P) \geq R(W_{\gamma,\eta}(P))$ where wiring acting on subsystems 2 and 3 is considered. Let us
fix $r$ and $s$ arbitrarily, and denote $\tau_{rs}(W_{\gamma,\eta}(P)) \equiv \tau_{rs}(P')$.
In analogy to the proof of the theorem \ref{thm:cost-lbound}, the case when for all $r's't'$ there is  $|\beta_{r's't'}(\tau_{rs}(P'))|\leq 2$ implies that $R(P')$ is zero (the box is local) \cite{Joshi-broadcasting} and the RHS of  (\ref{eq:robustness}) is not positive, hence the claimed inequality is satisfied. 

Let us consider now the non-trivial case
when there exist $r's'$ such that for all $t'$, there is $|\beta_{r's't'}(\tau_{rs}(P'))|>2$.
The box $\tau_{rs}(P')$ as $\tau_{rs}(P') = P^{\alpha}_{rst}$ is now described for $\alpha \in ({3\over 4},1]$ which fixes the value of $t\in\{0,1\}$.
Now, due to the observation \ref{obs:beta}  $r'=r,s'=s$, by choosing also $t' = t$:
\be
\beta_{rst}(P^{\alpha}_{rst}) > 2.
\ee

Then, from the equation (\ref{eq:r-beta}), there is:
\be
R(\tau_{rs}(P')) = \frac{\beta_{rst}(\tau_{rs}(P'))-2}{\beta_{rst}(\tau_{rs}(P'))+4}.
\ee
From the equation (\ref{eq:r-inv-t}) there is $R(\tau_{rs}(P')) = R(P')$ as robustness (like anti-robustness), is invariant under appropriate twirling: namely, if a box B has $\beta_{rst}(B) > 2$ then after $\tau_{rs}$,  $R(\tau_{rs}(B)) = R(B)$.

Finally, it is worth noticing that $\beta_{rst}(\tau_{rs}(P')) = \beta_{rst}(P')$ by observation \ref{obs:beta}. Since, as in the proof of theorem \ref{thm:cost-lbound}, $r,s,t$ are such, that: 
\be
{{\beta_{rst}(P') - 2}\over \beta_{rst}(P') + 4} \geq {\beta_{r''s''t''}(P') - 2\over \beta_{r''s''t''}(P') + 4}  
\ee
for any $r'',s'',t''\in\{0,1\}$, the LHS of the above inequality is the highest value of the RHS expression over $r'',s'',t''$. After maximization over $W$: 
\be
R_{XYZ}(P) \geq \max_{W} R(W(P)) \geq  \max_{\beta, W} \frac{\beta(W(P))-2}{\beta(W(P))+4},
\ee
as desired. $\Box$

{\it Proof of observation \ref{obs:res}}.
It is worth to consider a wiring $(x_2 = a_1, a_1'= a_2)$ performed on a box $P(a_1,a_2,a_3|x_1,x_2,x_3)$ leading to the box $P_1(a_1',a_3|x_1,x_3)$. The output of $P(a_1,a_2,a_3|x_1,x_2,x_3)$ can be locally changed, defining $\tilde{a}_1=a_1+a_1 x_1$. Now, the wiring $(x_2 = \tilde{a}_1, \tilde{a}_1'= a_2)$ leading to the resulting box $P_2(\tilde{a}_1',a_3|x_1,x_3)$ is considered. The same box $P_2(\tilde{a}_1',a_3|x_1,x_3)$ can be obtained by performing the wiring $(x_2 = a_1+a_1 x_1, a_1'= a_2)$ on the box $P(a_1,a_2,a_3|x_1,x_2,x_3)$. So the investigation of $(x_2 = a_1+a_1 x_1, a_1'= a_2)$ can be performed using $(x_2 = a_1, a_1'= a_2)$ and a locally relabeled box $P(\tilde{a}_1,a_2,a_3|x_1,x_2,x_3)$ with $\tilde{a}_1=a_1+a_1 x_1$ . In general, there is a correspondence between $(x_2 = a_1, a_1'= a_2)$ and all local relabellings $l(.)$ of the output $a_1$ which results in $(x_2 = l(a_1), a_1' =  \oplus_{ijk}\eta_{ijk}( l(a_1)^i x_1^j a_2^k) )$.$\Box$


\appendix


\bibliographystyle{apsrev}
\bibliography{wirbib}

\begin{table}[p!]
\begin{tabular}{ |l|l|l|l|l|l|} \hline
No.&$WN_{NNS}$&$a_1$'&No.&$WN_{NNS}$&$a_1$' \\ \hline 
1&3&$a_2$&19&$\frac{14}{5}$&$a_1 a_2+a_2 x_1+a_1 a_2 x_1$\\
2&3&$a_1+a_2$&20&$\frac{14}{5}$&$a_1 a_2+x_1+a_2 x_1+a_1 a_2 x_1$\\
3&3&$a_2+x_1$&21&$\frac{14}{5}$&$a_1 a_2+a_1 x_1+a_2 x_1+a_1 a_2 x_1$\\
4&3&$a_1+a_2+x_1$&22&$\frac{14}{5}$&$a_1 a_2+x_1+a_1 x_1+a_2 x_1+a_1 a_2 x_1$\\
5&$\frac{14}{5}$&$a_2+a_1 a_2 x_1$&23&$\frac{14}{5}$&$a_2+a_2 x_1+a_1 a_2 x_1$\\
6&$\frac{14}{5}$&$a_2+a_1 x_1+a_1 a_2 x_1$&24&$\frac{14}{5}$&$a_2+a_1 x_1+a_2 x_1+a_1 a_2 x_1$\\
7&$\frac{14}{5}$&$a_2+x_1+a_1 a_2 x_1$&25&$\frac{14}{5}$&$a_2+x_1+a_2 x_1+a_1 a_2 x_1$\\
8&$\frac{14}{5}$&$a_2+x_1+a_1 x_1+a_1 a_2 x_1$&26&$\frac{14}{5}$&$a_2+x_1+a_1 x_1+a_2 x_1+a_1 a_2 x_1$\\
9&$\frac{14}{5}$&$a_2+a_1 a_2+a_1 a_2 x_1$&27&$\frac{14}{5}$&$a_2+a_1 a_2+a_1 x_1+a_1 a_2 x_1$\\
10&$\frac{14}{5}$&$a_2+a_1 a_2+x_1+a_1 a_2 x_1$&28&$\frac{14}{5}$&$a_2+a_1 a_2+x_1+a_1 x_1+a_1 a_2 x_1$\\
11&$\frac{14}{5}$&$a_1+a_2+a_1 a_2 x_1$&29&$\frac{14}{5}$&$a_1+a_2+a_2 x_1+a_1 a_2 x_1$\\
12&$\frac{14}{5}$&$a_1+a_2+a_1 x_1+a_1 a_2 x_1$&30&$\frac{14}{5}$&$a_1+a_2+a_1 x_1+a_2 x_1+a_1 a_2 x_1$\\
13&$\frac{14}{5}$&$a_1+a_2+x_1+a_1 a_2 x_1$&31&$\frac{14}{5}$&$a_1+a_2+x_1+a_2 x_1+a_1 a_2 x_1$\\
14&$\frac{14}{5}$&$a_1+a_2+x_1+a_1 x_1+a_1 a_2 x_1$&32&$\frac{14}{5}$&$a_1+a_2+x_1+a_1 x_1+a_2 x_1+a_1 a_2 x_1$\\
15&$\frac{14}{5}$&$a_1+a_2+a_1 a_2+a_1 a_2 x_1$&33&$\frac{14}{5}$&$a_1+a_2+a_1 a_2+a_1 x_1+a_1 a_2 x_1$\\
16&$\frac{14}{5}$&$a_1+a_2+a_1 a_2+x_1+a_1 a_2 x_1$&34&$\frac{14}{5}$&$a_1+a_2+a_1 a_2+x_1+a_1 x_1+a_1 a_2 x_1$\\
17&$\frac{14}{5}$&$1+a_1 a_2+a_2 x_1+a_1 a_2 x_1$&35&$\frac{14}{5}$&$1+a_1 a_2+a_1 x_1+a_2 x_1+a_1 a_2 x_1$\\
18&$\frac{14}{5}$&$1+a_1 a_2+x_1+a_2 x_1+a_1 a_2 x_1$&36&$\frac{14}{5}$&$1+a_1 a_2+x_1+a_1 x_1+a_2 x_1+a_1 a_2 x_1$\\
\hline
\end{tabular}
\caption{The value of the WN for Wiring acting on subsystems 12 of boxes $P(a_1,a_2,a_3|x_1,x_2,x_3)$ belonging to the NNS correlations (namely, the maximal violation over the CHSH inequalities (\ref{eq:chshineq}) obtained using given wiring on the boxes belonging to the NNS class). Due to the observation \ref{obs:res} an input to the second subsystem is given by $x_2=a_1$. An output of the effective box $P(a_1',a_3|x_1,x_3)$ is given by $a_1'$.  }
\label{tab:nns}
\end{table}

\begin{table}[p!]
\begin{tabular}{ |l|l|l|l|l|l|} \hline
No.&$WN_{TTS}$&$a_1$'&No.&$WN_{TTS}$&$a_1$' \\ \hline 
1&3&$a_2$&23& $\frac{38}{13}$&$a_1 a_2+a_2 x_1+a_1 a_2 x_1$\\
2&3&$a_2+x_1$&24& $\frac{38}{13}$&$a_1 a_2+x_1+a_2 x_1+a_1 a_2 x_1$\\
3&3&$a_1+a_2 x_1$&25&$\frac{38}{13}$&$a_2+a_1 a_2 x_1$\\
4&3&$a_1+a_2$&26&$\frac{38}{13}$&$a_2+a_1 x_1+a_1 a_2 x_1$\\
5&3&$a_1+a_2+x_1$&27& $\frac{38}{13}$&$a_2+x_1+a_1 a_2 x_1$\\
6&3&$1+a_1+a_2 x_1$&28&$\frac{38}{13}$&$a_2+x_1+a_1 x_1+a_1 a_2 x_1$\\
7&3&$a_2+a_1 x_1+a_2 x_1$&29&$\frac{38}{13}$&$a_1 a_2+a_1 x_1+a_2 x_1+a_1 a_2 x_1$\\
8&3&$a_2+x_1+a_1 x_1+a_2 x_1$&30&$\frac{38}{13}$&$a_1 a_2+x_1+a_1 x_1+a_2 x_1+a_1 a_2 x_1$\\
9&3&$a_1+a_1 x_1+a_2 x_1$&31&$\frac{38}{13}$&$a_2+a_2 x_1+a_1 a_2 x_1$\\
10&3&$a_1+a_2+a_2 x_1$&32&$\frac{38}{13}$&$a_2+a_1 x_1+a_2 x_1+a_1 a_2 x_1$\\
11&3&$a_1+a_2+x_1+a_2 x_1$&33&$\frac{38}{13}$&$a_2+x_1+a_2 x_1+a_1 a_2 x_1$\\
12&3&$1+a_1+a_1 x_1+a_2 x_1$ &34&$\frac{38}{13}$&$a_2+x_1+a_1 x_1+a_2 x_1+a_1 a_2 x_1$\\
13&$\frac{38}{13}$&$a_2+a_1 a_2+a_1 a_2 x_1$&35&$\frac{38}{13}$&$a_2+a_1 a_2+a_1 x_1+a_1 a_2 x_1$\\14&$\frac{38}{13}$&$a_2+a_1 a_2+x_1+a_1 a_2 x_1$&36&$\frac{38}{13}$&$a_2+a_1 a_2+x_1+a_1 x_1+a_1 a_2 x_1$\\15&$\frac{38}{13}$&$a_1+a_2+a_1 a_2 x_1$&37&$\frac{38}{13}$&$a_1+a_2+a_2 x_1+a_1 a_2 x_1$\\16&$\frac{38}{13}$&$a_1+a_2+a_1 x_1+a_1 a_2 x_1$&38&$\frac{38}{13}$&$a_1+a_2+a_1 x_1+a_2 x_1+a_1 a_2 x_1$\\17&$\frac{38}{13}$&$a_1+a_2+x_1+a_1 a_2 x_1$&39&$\frac{38}{13}$&$a_1+a_2+x_1+a_2 x_1+a_1 a_2 x_1$\\18&$\frac{38}{13}$&$a_1+a_2+x_1+a_1 x_1+a_1 a_2 x_1$&40&$\frac{38}{13}$&$a_1+a_2+x_1+a_1 x_1+a_2 x_1+a_1 a_2 x_1$\\19&$\frac{38}{13}$&$a_1+a_2+a_1 a_2+a_1 a_2 x_1$&41&$\frac{38}{13}$&$a_1+a_2+a_1 a_2+a_1 x_1+a_1 a_2 x_1$\\20&$\frac{38}{13}$&$a_1+a_2+a_1 a_2+x_1+a_1 a_2 x_1$&42&$\frac{38}{13}$&$a_1+a_2+a_1 a_2+x_1+a_1 x_1+a_1 a_2 x_1$\\21&$\frac{38}{13}$&$1+a_1 a_2+a_2 x_1+a_1 a_2 x_1$&43&$\frac{38}{13}$&$1+a_1 a_2+a_1 x_1+a_2 x_1+a_1 a_2 x_1$\\22&$\frac{38}{13}$&$1+a_1 a_2+x_1+a_2 x_1+a_1 a_2 x_1$&44&$\frac{38}{13}$&$1+a_1 a_2+x_1+a_1 x_1+a_2 x_1+a_1 a_2 x_1$\\
\hline
\end{tabular}
\caption{The value of the WN for Wiring acting on subsystems 12 of boxes $P(a_1,a_2,a_3|x_1,x_2,x_3)$ belonging to the TTS correlations (namely, the maximal violation over the CHSH inequalities (\ref{eq:chshineq}) obtained using given wiring on boxes belonging to the TTS class). Due to the observation \ref{obs:res} an input to the second subsystem is given by $x_2=a_1$. An output of the effective box $P(a_1',a_3|x_1,x_3)$ is given by $a_1'$.}
\label{tab:tts}
\end{table}

\begin{table}[p]
\begin{tabular}{ |l|l|l|l|l|l|} \hline
No.&$WN_{NSS}$&$a_1$'&No.&$WN_{NSS}$&$a_1$' \\ \hline 
1&4&$a_2+a_1 x_1$&21&3&$a_1 a_2+a_2 x_1+a_1 a_2 x_1$\\ 
2&4&$a_1+a_2+a_1 x_1$&22&3&$a_1 a_2+x_1+a_2 x_1+a_1 a_2 x_1$\\ 
3&4&$a_2+x_1+a_1 x_1$&23&3&$a_1 a_2+a_1 x_1+a_2 x_1+a_1 a_2 x_1$\\
 4&4&$a_1+a_2+x_1+a_1 x_1$&24&3&$a_1 a_2+x_1+a_1 x_1+a_2 x_1+a_1 a_2 x_1$\\ 
5&3&$a_2$&25&3&$a_2+a_1 a_2 x_1$\\ 6&3&$a_2+a_2 x_1+a_1 a_2 x_1$&26&3&$a_2+a_1 x_1+a_1 a_2 x_1$\\
 7&3&$a_2+a_1 x_1+a_2 x_1+a_1 a_2 x_1$&27&3&$a_2+x_1$\\
 8&3&$a_2+x_1+a_1 a_2 x_1$&28&3&$a_2+x_1+a_2 x_1+a_1 a_2 x_1$\\ 
9&3&$a_2+x_1+a_1 x_1+a_1 a_2 x_1$&29&3&$a_2+x_1+a_1 x_1+a_2 x_1+a_1 a_2 x_1$\\ 
10&3&$a_2+a_1 a_2+a_1 a_2 x_1$&30&3&$a_2+a_1 a_2+a_1 x_1+a_1 a_2 x_1$\\ 
11&3&$a_2+a_1 a_2+x_1+a_1 a_2 x_1$&31&3&$a_2+a_1 a_2+x_1+a_1 x_1+a_1 a_2 x_1$\\ 
12&3&$a_1+a_2$&32&3&$a_1+a_2+a_1 a_2 x_1$\\ 
13&3&$a_1+a_2+a_2 x_1+a_1 a_2 x_1$&33&3&$a_1+a_2+a_1 x_1+a_1 a_2 x_1$\\ 
14&3&$a_1+a_2+a_1 x_1+a_2 x_1+a_1 a_2 x_1$&34&3&$a_1+a_2+x_1$\\ 
15&3&$a_1+a_2+x_1+a_1 a_2 x_1$&35&3&$a_1+a_2+x_1+a_2 x_1+a_1 a_2 x_1$\\ 
16&3&$a_1+a_2+x_1+a_1 x_1+a_1 a_2 x_1$&36&3&$a_1+a_2+x_1+a_1 x_1+a_2 x_1+a_1 a_2 x_1$\\ 
17&3&$a_1+a_2+a_1 a_2+a_1 a_2 x_1$&37&3&$a_1+a_2+a_1 a_2+a_1 x_1+a_1 a_2 x_1$\\ 
18&3&$a_1+a_2+a_1 a_2+x_1+a_1 a_2 x_1$&38&3&$a_1+a_2+a_1 a_2+x_1+a_1 x_1+a_1 a_2 x_1$\\ 
19&3&$1+a_1 a_2+a_2 x_1+a_1 a_2 x_1$&39&3&$1+a_1 a_2+a_1 x_1+a_2 x_1+a_1 a_2 x_1$\\ 
20&3&$1+a_1 a_2+x_1+a_2 x_1+a_1 a_2 x_1$&40&3&$1+a_1 a_2+x_1+a_1 x_1+a_2 x_1+a_1 a_2 x_1$\\  \hline
\end{tabular}
\caption{The value of the WN for Wiring acting on subsystems 12 of boxes $P(a_1,a_2,a_3|x_1,x_2,x_3)$ belonging to the NSS correlations (namely, maximal violation over the CHSH inequalities (\ref{eq:chshineq}) obtained using given wiring on the boxes belonging to NSS class). Due to observation \ref{obs:res} an input to the second subsystem is given by $x_2=a_1$. An output of the effective box $P(a_1',a_3|x_1,x_3)$ is given by $a_1'$.}
\label{tab:nss}
\end{table}

\begin{table}[p]
\begin{tabular}{ |l|l|l|l|l|l|} \hline
No.&$WN_{TSS}$&$a_1$'&No.&$WN_{TSS}$&$a_1$' \\ \hline 
1&4&$a_2+a_1 x_1$&25&3&$a_1 a_2+a_2 x_1+a_1 a_2 x_1$\\
2&4&$a_1+a_2+a_1 x_1$&26&3&$a_1 a_2+x_1+a_2 x_1+a_1 a_2 x_1$\\ 
3&4&$a_2+x_1+a_1 x_1$&27&3&$a_1 a_2+a_1 x_1+a_2 x_1+a_1 a_2 x_1$\\ 
4&4&$a_1+a_2+x_1+a_1 x_1$&28&3&$a_1 a_2+x_1+a_1 x_1+a_2 x_1+a_1 a_2 x_1$\\ 
5&3&$a_2$&29&3&$a_2+a_1 a_2 x_1$\\
6&3&$a_2+a_2 x_1+a_1 a_2 x_1$&30&3&$a_2+a_1 x_1+a_1 a_2 x_1$\\
7&3&$a_2+a_1 x_1+a_2 x_1$&31&3&$a_2+a_1 x_1+a_2 x_1+a_1 a_2 x_1$\\
8&3&$a_2+x_1$&32&3&$a_2+x_1+a_1 a_2 x_1$\\ 
9&3&$a_2+x_1+a_2 x_1+a_1 a_2 x_1$&33&3&$a_2+x_1+a_1 x_1+a_1 a_2 x_1$\\
10&3&$a_2+x_1+a_1 x_1+a_2 x_1$&34&3&$a_2+x_1+a_1 x_1+a_2 x_1+a_1 a_2 x_1$\\ 
11&3&$a_2+a_1 a_2+a_1 a_2 x_1$&35&3&$a_2+a_1 a_2+a_1 x_1+a_1 a_2 x_1$\\ 
12&3&$a_2+a_1 a_2+x_1+a_1 a_2 x_1$&36&3&$a_2+a_1 a_2+x_1+a_1 x_1+a_1 a_2 x_1$\\ 
13&3&$a_1+a_2 x_1$&37&3&$a_1+a_1 x_1+a_2 x_1$\\ 14&3&$a_1+a_2$&38&3&$a_1+a_2+a_1 a_2 x_1$\\ 
15&3&$a_1+a_2+a_2 x_1$&39&3&$a_1+a_2+a_2 x_1+a_1 a_2 x_1$\\
16&3&$a_1+a_2+a_1 x_1+a_1 a_2 x_1$&40&3&$a_1+a_2+a_1 x_1+a_2 x_1+a_1 a_2 x_1$\\ 
17&3&$a_1+a_2+x_1$&41&3&$a_1+a_2+x_1+a_1 a_2 x_1$\\
18&3&$a_1+a_2+x_1+a_2 x_1$&42&3&$a_1+a_2+x_1+a_2 x_1+a_1 a_2 x_1$\\ 
19&3&$a_1+a_2+x_1+a_1 x_1+a_1 a_2 x_1$&43&3&$a_1+a_2+x_1+a_1 x_1+a_2 x_1+a_1 a_2 x_1$\\ 
20&3&$a_1+a_2+a_1 a_2+a_1 a_2 x_1$&44&3&$a_1+a_2+a_1 a_2+a_1 x_1+a_1 a_2 x_1$\\ 
21&3&$a_1+a_2+a_1 a_2+x_1+a_1 a_2 x_1$&45&3&$a_1+a_2+a_1 a_2+x_1+a_1 x_1+a_1 a_2 x_1$\\ 
22&3&$1+a_1 a_2+a_2 x_1+a_1 a_2 x_1$&46&3&$1+a_1 a_2+a_1 x_1+a_2 x_1+a_1 a_2 x_1$\\ 
23&3&$1+a_1 a_2+x_1+a_2 x_1+a_1 a_2 x_1$&47&3&$1+a_1 a_2+x_1+a_1 x_1+a_2 x_1+a_1 a_2 x_1$\\
24&3&$1+a_1+a_2 x_1$&48&3&$1+a_1+a_1 x_1+a_2 x_1$\\ 
\hline
\end{tabular}
\caption{The value of the WN for Wiring acting on subsystems 12 of boxes $P(a_1,a_2,a_3|x_1,x_2,x_3)$ belonging to the TSS correlations (namely, the maximal violation over the CHSH inequalities (\ref{eq:chshineq}) obtained using given wiring on the boxes belonging to TSS class). Due to the observation \ref{obs:res} an input to the second subsystem is given by $x_2=a_1$. An output of the effective box $P(a_1',a_3|x_1,x_3)$ is given by $a_1'$.}
\label{tab:tss}
\end{table}

\end{document}